\documentclass[12pt]{article}
\usepackage[english]{babel}
\usepackage[utf8]{inputenc} 
\usepackage{lmodern}
\usepackage[normalem]{ulem}
\usepackage{amsmath, amsthm, amssymb}
\usepackage{geometry}
\usepackage{enumerate}
\usepackage{emptypage}
\usepackage{multirow}
\usepackage{xcolor} % in the preamble
\definecolor{darkred}{RGB}{139,0,0} % define a dark red color
\usepackage[title]{appendix}
\usepackage{cite}
\usepackage[colorlinks=true, linkcolor=blue, citecolor=blue, urlcolor=darkred]{hyperref}
\usepackage{bookmark}
\usepackage{graphicx}
\usepackage{comment}
\usepackage{placeins}
\usepackage{tikz-feynman}
\tikzfeynmanset{compat=1.1.0}
\usepackage{caption}
\usepackage{subfig}

\usepackage{hyperref}
\usepackage{amsfonts}
\usepackage{makeidx}
\usepackage{bbm}
\usepackage{slashed}
\usepackage{booktabs}
\usepackage{braket}
\usepackage[symbol]{footmisc}
\usepackage{mathtools}
\usepackage{float}
\usepackage{diagbox}

\usepackage{bold-extra}

\def\dd{\displaystyle}

\numberwithin{equation}{section}

\newcommand{\re}{\text{Re}\,}
\newcommand{\im}{\text{Im}\,}

\newcommand{\diag}{\text{diag}}

\newcommand{\beq}{\begin{equation}}
\newcommand{\eeq}{\end{equation}}

\begin{document}
\vspace{0.1cm}
\begin{center}
{\LARGE \bf On quark-lepton mixing and the leptonic CP violation}\\[2ex]
\vspace{1cm}
{Alessio Giarnetti\footnote[1]{e-mail address: \href{mailto:alessio.giarnetti@uniroma3.it}{\color{blue}alessio.giarnetti@uniroma3.it}}}\,,
{Simone Marciano\footnote[2]{e-mail address: \href{mailto:simone.marciano@uniroma3.it}{\color{blue}simone.marciano@uniroma3.it}}}\,,
{Davide Meloni\footnote[3]{e-mail address: \href{mailto:davide.meloni@uniroma3.it}{\color{blue}davide.meloni@uniroma3.it}}} \\[5mm]

{\textit{Dipartimento di Matematica e Fisica, 
Universit\`a di Roma Tre}}\\
{\textit{INFN Sezione di Roma Tre, Via della Vasca Navale 84, 00146, Roma, Italy}}
\vspace{16pt}

\begin{center}{\bf Abstract}
\begin{quote}
In the absence of a Grand Unified Theory framework, connecting the values of the mixing parameters in the quark and lepton sector is a difficult task, unless one introduces ad-hoc relations among the matrices that diagonalize such different kinds of fermions. In this paper, we discuss in detail the possibility that the PMNS matrix is given by the product $U_{PMNS}= V_{CKM}^\star\,T^\star$, where $T$ comes from the diagonalization of a see-saw like mass matrix that can be of a Bimaximal (BM), Tri-Bimaximal (TBM) and Golden Ratio (GR) form, and identify the leading corrections to such patterns that allow a good fit to the leptonic mixing matrix as well as to the CP phase. We also show that the modified versions of BM, TBM and GR can easily  accommodate the solar and atmospheric mass differences.

\end{quote}
\end{center}
\end{center}

%%%%%%%%%%%%%%%%%%%%%%%% 1.  INTRODUCTION   %%%%%%%%%%%%%%%%%%%%%%%%%%%%%%
%

\renewcommand{\thefootnote}{\arabic{footnote}}
\setcounter{footnote}{0}
\setcounter{tocdepth}{1}
{\let\clearpage\relax \tableofcontents}

\section{Introduction}
In the last years, neutrino experiments confirmed that neutrinos oscillate and measured with a great precision the values of the mixing angles. Some neutrino oscillation properties are still unknown/not really clear (as, for example, whether CP violation exists in the lepton sector or whether the mass hierarchy is of normal or inverted type) but the emerging picture is quite intriguing: differently from the mixing angles in the quark sector, described by an almost diagonal matrix $V_{CKM}$, neutrino mixing is dictated by two large and one small angle, thus making the $U_{PMNS}$ a matrix with large entries, except for the $(13)$ element. In spite of this huge discrepancy (that has been dubbed as {\it the flavor problem}), the current numerical values of fermion mixings seem to be inextricably wedged into well-defined relations \cite{Minakata_2004} which, using the standard parametrization of mixing matrices, are summarized as follows: 
\begin{equation}
\label{qlcrelations}
\theta_{12}^{PMNS} +\theta_{12}^{CKM} \sim \pi/4 \,,\qquad  \theta_{23}^{PMNS} +\theta_{23}^{CKM} \sim \pi/4\,.
\end{equation}
The previous  structure, which is presumed to exist behind such 
empirical relations, is known as {\it quark-lepton complementarity (QLC)} and, while being appealing from a theoretical and phenomenological point of view, does not give any clue on which kind of symmetry could be responsible for them.

The usual answer to this problem is grand unification (GUT) in which quarks and leptons are unified into the same multiplets
\cite{Georgi:1979df,Raidal_2004,Frampton_2005,Antusch_2005,Xing_2005,Datta:2005ci,Patel:2010hr}; on the other hand, in non-GUT scenarios, one is somehow forced to input the CKM (PMNS) matrix into the relations that define the PMNS (CKM).  Several authors have explored such a possibility \cite{Minakata_2004,PICARIELLO_2007,Chauhan_2007,Harada_2006,Zhukovsky:2019eoy,Zhang:2012pv,Zhang:2012zh,Barranco:2010we}
and discussed the observable consequences of scenarios leading to QLC \cite{Cheung_2005,Hochmuth_2007,Plentinger:2007px,Kang:2011tv,Ke:2014hxa}, including the effect of the RGE running on the stability of eq.(\ref{qlcrelations}) \cite{Ferrandis:2004vp,Kang_2005,Schmidt:2006rb,Dighe_2006}. An extension of eq.(\ref{qlcrelations}) to the $(13)$ sector results in a complete failure, as the sum $\theta_{13}^{PMNS} +\theta_{13}^{CKM} \sim 10^\circ$; thus, it is necessary to find a new connection between neutrinos and quarks that involves the reactor angle. 
The most promising suggestion is, once again, GUT-inspired and reads:
\begin{equation}
\label{qlcrelations2}
\theta_{13}^{PMNS} = \alpha \,\theta_{12}^{CKM}\,,
\end{equation}
where $\alpha$ can be any ${\cal O}(1)$ number 
\cite{Ferrandis:2004mq,Chauhan_2007,Meloni:2011fx,Harada:2013aja,Sharma:2015dqi}.
One possibility to recover eq.(\ref{qlcrelations2}) is to assume that the mixing matrices are related through:
\begin{eqnarray}
\label{qlcnoi}
U_{PMNS} &\sim& V_{CKM}\,T\,,
\end{eqnarray}
where $T$ is an appropriate unitary matrix that we parametrize as the product of three sub-roations:
\begin{equation}
\label{defT}
T \equiv U_{23} U_{13} U_{12}    \,.
\end{equation}
In this paper we want to elaborate more on eq.(\ref{qlcnoi}), finding the exact theoretical relation among $U_{PMNS}$ and $V_{CKM}$ allowed by specific ansatz on the diagonalization procedure of the fermion mass matrices. This involves the determination of the matrix $T$; by assuming an initial form for $T$ of Bimaximal (BM), Tri-Bimaximal (TBM) \cite{Harrison:2002er} and Golden Ratio (GR) type \cite{Feruglio_2011}, we compute in a systematic way all relevant corrections that allow to reproduce the neutrino mixing angles as well as the Jarlskog invariant \cite{PhysRevLett.55.1039}. Instead of performing an overall fit involving general perturbations of BM, TBM and GR mixings, we preferred to introduce three different corrections, one for each $U_{ij}$ quoted in eq.(\ref{defT}), and study the prediction of mixing parameters determined by each of them.
In this way, we are able to keep track of the relevant source of deviations from the initial form for $T$ that allows a good fit to the experimental data.
We find that a complex parameter $u$ is
needed in the $U_{13}$ rotation to increase the amount of leptonic
CP violation up to the current experimental values while a simple real correction in the $(12)$ plane is mandatory to account for the solar angle. Finally, deviation to maximality for the atmospheric angle can be accounted by a real shift $\omega$ in the (23) sector. 
In addition to mixing parameters, the newly found corrections are also compatible with the solar and atmospheric mass differences to a high degree of precision.

The paper is organized as follows: in Sect.(\ref{corrmix}) we discuss all the above-mentioned corrections in detail, showing how to include them in a perturbative approach to the determination of the mixing parameters; in Sect.(\ref{neumasses}) we show how to reproduce the experimental mass differences within our framework for all perturbed mixing patterns; finally, Sect.(\ref{conclusions}) is devoted to our conclusions.
We close the paper with the Appendix where we report  the expressions of the mixing parameters up to ${\cal O}(\lambda^3)$.

\section{Corrections to BM, TBM and GR}
\label{corrmix}

\subsection{Notation}
Let us first fix our notation; we are working in the left-right (LR) basis and, with no loss of generality, we assume diagonal heavy right-handed neutrinos $M_R = M_R^{diag}$ and diagonal charged leptons $M_\ell = M_\ell^{diag}$.
The diagonalization of the Dirac neutrino mass is achieved through $W_L^\dagger m_{\nu_D} U_R = m^{diag}_{\nu_D}$, so that the hermitean matrix $ m_{\nu_D} m_{\nu_D}^\dagger$ is such that $W_L^\dagger  m_{\nu_D} m_{\nu_D}^\dagger W_L = (m^{diag}_{\nu_D})^2 $, where the eigenvalues of $(m^{diag}_{\nu_D})^2$ are real and non-negative, and the columns of $W_L$ are the eigenvectors of the $ m_{\nu_D} m_{\nu_D}^\dagger$ matrix. Applying the see-saw formula in the LR basis, we get:
\begin{eqnarray}
m_\nu &=& -m_{\nu_D}\, (M_R^{diag})^{-1} \,m_{\nu_D}^T \nonumber \\ 
&=& W_L    \,m^{diag}_{\nu_D} \,U_R^\dagger \,(M_R^{diag})^{-1}\, U_R^\star \,
\label{eq1}
m^{diag}_{\nu_D}\,W_L^T\,.
\end{eqnarray}
At this point, the matrix $m_0 = m^{diag}_{\nu_D} \,U_R^\dagger \,(M_R^{diag})^{-1}\, U_R^\star \,
m^{diag}_{\nu_D}$ is a complex symmetric matrix and, thus, it can be diagonalized by an unitary matrix $T$ such that: 
\begin{eqnarray}
\label{m0}
m_0 &=& T\,S\,T^T\,,
\end{eqnarray}
where $S$ is a diagonal matrix with, in principle, complex entries.
%$\{\sigma_1,\sigma_2,\sigma_3\}$ with positive entries such that $\sigma_1 > \sigma_2> \sigma_3$. 
Thus, for the light neutrino mass we have the following decomposition:
\begin{eqnarray}
m_\nu &=& - W_L\,T\,S\,T^T\,W_L^T\,.
\end{eqnarray}
To get the proper structure of $U_{PMNS}$, we assume a neutrino change of basis of the following type:
\begin{eqnarray}
\nu^\prime &=& U_{PMNS}\,\nu\,,
\end{eqnarray}
where the mass eigenstate are those indicated with $\nu$. 
At the Lagrangian level, the symmetric mass term, in the basis of interaction eigenstates, is as follows:
\begin{eqnarray}
(\nu^T)^\prime\,m_\nu \nu^\prime &=& \nu^T\,U_{PMNS}^T\,m_\nu \, U_{PMNS}\,\nu \equiv \nu^T\,m_\nu^{diag}\,\nu
\end{eqnarray}
so that: 
\begin{eqnarray}
m_\nu &=& U_{PMNS}^\star\,m_\nu^{diag}\,U_{PMNS}^\dagger\,,
\end{eqnarray}
and we can identify:
\begin{eqnarray}
\label{relUPMNS}
U_{PMNS} &=& W_L^\star\,T^\star\,
\end{eqnarray}
and 
\begin{eqnarray}
m_\nu^{diag} &=& S\,.
\end{eqnarray}
In the following we will assume $W_L \equiv V_{CKM}$, whose structure in the Wolfenstein paramterization is reported below:
\begin{equation}
\label{pmns_full}
V_{CKM} =
\begin{pmatrix}
1-\lambda^2/2 & \lambda  & A \lambda^3 (-i \eta + \rho) \\
-\lambda & 1-\lambda^2/2
 &A \lambda^2 \\
A \lambda^3 (1 - i \eta - \rho) & -A \lambda^2 & 1
\end{pmatrix}
\,.
\end{equation}
The values of the $V_{CKM}$ parameters used in our simulations are provided in Tab.(\ref{nufit_table2}).
\begin{table}[h]
\centering
\renewcommand{\arraystretch}{1.2}
\begin{tabular}{l c c} 
\toprule
Parameter$\qquad\qquad$ & \multicolumn{2}{c}{Best-fit value and $1\sigma$ range} \\ 
\midrule
$\lambda$ & $0.2251\pm 0.0008$ \\
$A$ & $0.828\pm  0.01$ \\
$\eta$ & $0.355\pm 0.009$ \\
$\rho$ & $0.164\pm 0.009$ \\
\bottomrule
\end{tabular}
\caption{\small{\it{Best-fit value and $1\sigma$ range of the $V_{CKM}$ parameters, from \cite{utfit}.}
}}
\label{nufit_table2}
\end{table}
For the $T$ matrix, instead, one can in principle assume an exact Tri-Bimaximal mixing (TBM), Bimaximal mixing (BM) or Golden Ratio (GR) forms:
\begin{equation}
U_{BM}= \left(
\begin{array}{ccc}
\dd\frac{1}{\sqrt 2}&\dd-\frac{1}{\sqrt 2}&0\\
\dd\frac{1}{2}&\dd\frac{1}{2}&\dd\frac{1}{\sqrt 2}\\
-\dd\frac{1}{2}&-\dd\frac{1}{2}&\dd\frac{1}{\sqrt 2}
\end{array}
\right) ~~~
U_{TBM}=
	\left(
	\begin{array}{ccc}
		\sqrt{\frac{2}{3}}  & \frac{1}{\sqrt{3}} & 0 \\
		-\frac{1}{\sqrt{6}} & \frac{1}{\sqrt{3}} & \frac{1}{\sqrt{2}} \\
		\frac{1}{\sqrt{6}} & -\frac{1}{\sqrt{3}} & \frac{1}{\sqrt{2}} \\
	\end{array}
	\right) ~~~ 
U_{GR}= \left(
\begin{array}{ccc}
c_{12}&s_{12}&0\\\
\dd\frac{s_{12}}{\sqrt 2}&-\dd\frac{ c_{12}}{\sqrt 2}&\dd\frac{1}{\sqrt 2}\\
\dd\frac{s_{12}}{\sqrt 2}&-\dd\frac{ c_{12}}{\sqrt 2}&-\dd\frac{1}{\sqrt 2}
\end{array}
\right)
\label{pred}
\end{equation}
where $c_{12}=\cos \theta_{12}, s_{12}=\sin\theta_{12}$ and $\tan\theta_{12}=1/\phi$, with $\phi =  (1 + \sqrt{5})/2$. 
However, it turns out that the $U_{PMNS}$ implied by them is unsatisfactory in the predicted values of the mixing angles and Jarlskog invariant $J_{\text{CP}}$, for which we use the following expression:
\begin{eqnarray}
J_{\text CP} = \im\left[(U_{PMNS})_{11}(U_{PMNS})_{12}^*(U_{PMNS})_{21}^*(U_{PMNS})_{22}\right]\,.
\end{eqnarray}
We have summarized the situation in Tab.(\ref{zeroth}) where, for each mixing pattern, we have reported  the perturbative prediction on $\sin(\theta_{13}), \tan(\theta_{12}), \tan(\theta_{23})$ (up to ${\cal O}(\lambda$)) and $J_{\text{CP}}$ (up to ${\cal O}(\lambda^3$)). In the last column we have computed the {\it distance} $\Delta$ between such predictions and the current experimental values for a Normal Ordering (NO) of the neutrino masses\footnote{For our purposes, it is enough to consider the normal hierarchy only, as the only significant difference  with respect to the inverted ordering case is  a slight preference for the opposite $\theta_{23}$ octant.}, reported in Tab.(\ref{nufit_table}). Such a {\it distance} is computed according the following formula:
\begin{eqnarray}
\Delta=\Sigma_{i=1}^3  \left[\frac{P_i-B_i}{\sigma_i}\right]^2\,,
%\frac{\left[J-J^{bf}\right]^2}{\sigma_{J}^2}   +  \frac{\left[\sin(\theta_{13})-\sin(\theta_{13})^{bf}\right]^2}{\sigma_{\sin(\theta_{13})}^2} +  \frac{\left[\tan(\theta_{23})-\tan(\theta_{23})^{bf}\right]^2}{\sigma_{\tan(\theta_{23})}^2} + 
% \frac{\left[\tan(\theta_{12})-\tan(\theta_{12})^{bf}\right]^2}{\sigma_{\tan(\theta_{12})}^2} \,.\nonumber \\
\label{delta}
\end{eqnarray}
where $\vec{P}$ is a vector of parameters $\vec{P}=\left[\tan(\theta_{12}),\tan(\theta_{13}),\tan(\theta_{23}),J_{\text{CP}}\right]$ as predicted by TBM, BM and GR (see Tab.(\ref{zeroth})),  $\vec{\sigma}$ are the related $1\sigma$ errors and  $\vec{B}$ contains the best-fit values of Tab.(\ref{nufit_table}), $\vec{B}=\left[\tan^{bf} (\theta_{12}),\tan^{bf}(\theta_{13}),\tan^{bf}(\theta_{23}),J^{bf}_{\text{CP}}\right]$. 
$\Delta$ allows us  to estimate how far a given texture is from the current values of the mixing parameters.
\begin{table}[t]
\centering
\renewcommand{\arraystretch}{1.2}
\begin{tabular}{l c c} 
\toprule
Parameter$\qquad\qquad$ &Best-fit value and $1\sigma$ range \\ 
\midrule
%
%& \textbf{NO} & \textbf{IO} \\

%
$r \equiv \Delta m^2_\text{sol}/|\Delta m^2_\text{atm}|$ & $0.0295\pm0.0008$ \\
$\tan(\theta_{12})$ & $0.666 \pm 0.019$  \\
$\sin(\theta_{13})$ & $0.149\pm 0.002$  \\
$\tan(\theta_{23})$ & $0.912\pm 0.035$  \\
$J_{\text{CP}}$ & $-0.027\pm 0.010$  \\
\bottomrule
\end{tabular}
\caption{\it{Neutrino observables and their 1$\sigma$ ranges as derived from NuFIT 5.3 \cite{Esteban:2020cvm,nufit}, using the dataset with SK atmospheric data \cite{Super-Kamiokande:2017yvm}.  For the  extraction of  the best-fit value and $1\sigma$ uncertainty of the Jarlskog invariant, we refer to its one-dimensional $\chi^2$ projection from NuFIT 5.3.}
}
\label{nufit_table}
\end{table}

\begin{figure}[h!]
  \begin{center}
    \includegraphics[scale=0.6]{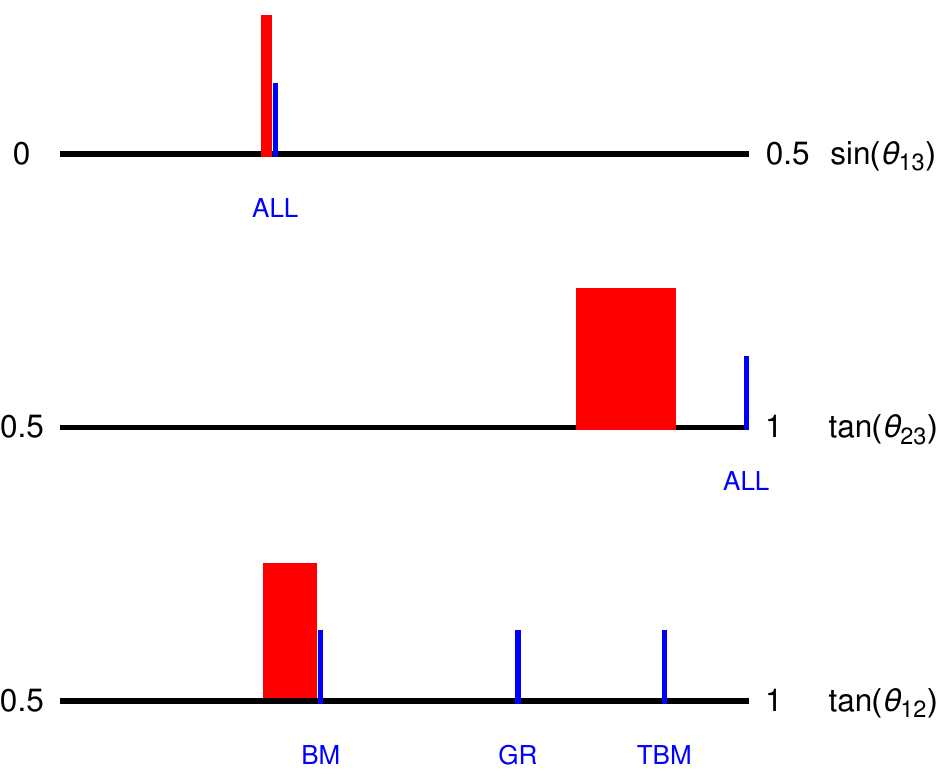}
    \caption{\small{\it  Current 1$\sigma$ experimental spread on $\tan(\theta_{12}),\tan(\theta_{23})$ and $\sin(\theta_{13})$ (red rectangles) and the predictions derived from $U_{PMNS}=V_{CKM}^\star T^\star$, where $T$ can be TBM, BM and GR mixing patterns.}}
    \label{th12bar}
  \end{center}
  \end{figure}

\begin{table}[h]
\centering
\renewcommand{\arraystretch}{1.2}
\begin{tabular}{l c c c c c} 
\toprule
$T$ & $\sin(\theta_{13})$ & $\tan(\theta_{12})$ & $\tan(\theta_{23})$ & $J_{\text{CP}}$ & $\Delta$\\ 
\midrule
$U_{TBM}$ & $\frac{\lambda}{\sqrt{2}}$ &$\frac{1}{\sqrt{2}}+\frac{3\lambda}{2\sqrt{2}}$ & 1&  $-\frac{1}{6} A \eta \lambda^3$  & 2715\\
$U_{BM}$ &$\frac{\lambda}{\sqrt{2}}$&$1-\sqrt{2} \lambda$ & 1& $\frac{1}{4\sqrt{2}} A \eta \lambda^3$ & 2500\\
$U_{GR}$ & $\frac{\lambda}{\sqrt{2}}$ & $\frac{2\sqrt{5}}{5+\sqrt{5}}+\frac{5\sqrt{2}}{5+\sqrt{5}}\lambda$ & 1&  $-\frac{1}{2\sqrt{10}} A \eta \lambda^3$ & 2580\\
\bottomrule
\end{tabular}
\caption{\it{Perturbative predictions on $\sin(\theta_{13}), \tan(\theta_{12}), \tan(\theta_{23})$ (up to ${\cal O}(\lambda))$ and $J_{\text{CP}}$ (up to ${\cal O}(\lambda^3))$ as obtained from the ansatz $U_{PMNS}=V_{CKM}^\star T^\star$, where $T$ can be TBM, BM and GR mixing patterns. In the last column we report the values of the variable $\Delta$ defined in eq.(\ref{delta}).}
}
\label{zeroth}
\end{table}
While all patterns predict maximal $(23)$ mixing and the same $\sin(\theta_{13})$, the differences come from $J_{\text{CP}}$ (strongly suppressed for all patterns) and from the solar sector; in particular, for the latter the BM mixing results in a better agreement with the current experimental value than TBM and GR, as evident by the smaller $\Delta$. The predictions in Tab.(\ref{zeroth})  are also reported in Fig.(\ref{th12bar}), together with  their 1$\sigma$ experimental spread (red rectangles)\footnote{We do not report the spread of $J_{\text{CP}}$ as, for any patters, its absolute value is around two orders of magnitude smaller than the experimental best-fit.}. 
From this we learn that, after the shifts of ${\cal O}(\lambda)$ provided by $V_{CKM}$, negative corrections are needed for all patterns to jump into the 1$\sigma$ allowed range for all mixing angles. It is worth to mention that, if 3$\sigma$ allowed ranges for the atmospheric mixing angle and the Jarlskog invariant are taken into account, the BM scenario is compatible with experimental data. Indeed, both $\sin\theta_{23}\sim1$ and $J_{\text{CP}}\sim0$ are not yet excluded by neutrino experiments \cite{nufit}. 

In the next section we will analyze, in a systematic way, which corrections of $U_{ij}$ in eq.(\ref{defT}) are the most appropriate to better fit the neutrino mixing parameters.  

\subsection{Corrections from the $(13)$-sector to BM, TBM and GR}

We start our analysis by studying in detail the correction to the standard patterns from the $(13)$-sector. 
The main idea is that, given the absence of any CP phase in (\ref{pred}),  eq.(\ref{relUPMNS}) implies
a very low CP violation in the lepton sector \cite{Farzan:2006vj}, of the order of ${\cal O}(\lambda^3)$ and proportional to $\eta$ as shown from the expressions of $J_{\text{CP}}$ in Tab.(\ref{nufit_table}). Thus, to allow for a larger CP violation, which seems to be preferred by recent oscillation results, new sources of symmetry violation are needed.
Assuming for $T$ the decomposition as in eq.(\ref{defT}), larger CP violation can be generated by slightly shifting the $(13)$-rotation from the identity; to this aim, we introduce a complex parameter $u$ \cite{Altarelli:2004za} such that $|u| \ll 1$ and we rescale it by one power of the Cabibbo angle $\lambda$. This also implies that the rescaled $|u| \sim {\cal O}(1)$. 
Thus, the $(13)$-rotation has the following structure:
\begin{equation}
\label{notpmns2}
U_{13} =
\begin{pmatrix}
1 - \frac{\lambda^2}{2} |u|^2 &0  & u \lambda\\
0 & 1 & 0 \\
- u^* \lambda& 0 &1 - \frac{\lambda^2}{2} |u|^2
\end{pmatrix}
\,.
\end{equation}
To construct completely the matrix $T$, we need to specify the rotations in the other two sectors, the $(12)$ and $(23)$-rotations.  In order to contemplate the BM, TBM and GR mixings simultaneously, we leave unspecified the rotation in the $(12)$-sector and, since the sign of such a  rotation is not fixed a priori, we leave it as free, encoding this uncertainty into the parameter $\sigma$, that can assume values $\pm 1$. At this stage, the rotation in the $(23)$-sector is maximal (so, from our ansatz, we expect all deviations to $\theta_{23}$ coming from $V_{CKM}$, see below). 
Thus, we have:
\begin{equation}
\label{notpmns}
U_{23} =
\begin{pmatrix}
1 &0  & 0\\
0 & \frac{1}{\sqrt{2}}
 & \frac{1}{\sqrt{2}} \\
0 &
-\frac{1}{\sqrt{2}} & \frac{1}{\sqrt{2}} 
\end{pmatrix}
\,,\qquad 
U_{12} =
\begin{pmatrix}
\tilde{c}_{12} &  \sigma \tilde{s}_{12}  & 0\\
- \sigma \tilde{s}_{12} & \tilde{c}_{12} & 0 \\
0& 0 &1 
\end{pmatrix}\,,
\end{equation}
where $\tilde{c}_{12}\equiv\cos(\tilde{\theta}_{12}),\tilde{s}_{12}\equiv\sin(\tilde{\theta}_{12})$ are the cosinus and sinus functions of a rotation in the (12)-sector (not to be confused with the usual solar angle).
This, in turn, implies the following structure of the $T$ matrix:
\begin{equation}
\label{notpmns3}
T \equiv U_{23} U_{13} U_{12} =
\begin{pmatrix}
\tilde{c}_{12} \left(1 - \frac{\lambda^2}{2} |u|^2\right)& \sigma \tilde{s}_{12} \left(1 - \frac{\lambda^2}{2} |u|^2\right) & u \lambda \\
-(\sigma\tilde{s}_{12} + \tilde{c}_{12} u^*\lambda )/\sqrt{2} & (\tilde{c}_{12} - \sigma \tilde{s}_{12} u^* \lambda )/\sqrt{2}
 & \left(1 - \frac{\lambda^2}{2} |u|^2\right)/\sqrt{2} \\
(\sigma \tilde{s}_{12} - \tilde{c}_{12} u^*\lambda)/\sqrt{2} &
-(\tilde{c}_{12} + \sigma \tilde{s}_{12} u^*\lambda)/\sqrt{2} & \left(1 - \frac{\lambda^2}{2} |u|^2\right)/\sqrt{2} 
\end{pmatrix}
\,.
\end{equation}
Notice that unitarity is fully respected up to ${\cal O}(\lambda^3)$.
With our parametrization, the relevant patterns are recovered once we fix $u=0$ (for all of them) and  
$\tilde{s}_{12}=1/\sqrt{3},\tilde{s}_{12}=1/\sqrt{2}$ and $\tilde{s}_{12}^2=2/(5 + \sqrt{5})$ for TBM, BM and GR,  respectively (at this stage, the value of $\sigma$ is irrelevant).
For the Jarlskog invariant $J_{\text CP}$, up to ${\cal O}(\lambda^3)$, we get the expression as below:
%\begin{eqnarray}
%J&=& \frac{1}{4}\, \im{(u)} \,\sin^2(2 \tilde{\theta}_{12}) + \frac{\lambda}{2\sqrt{2}} \,
%\im{(u)} \left[\cos(2 \tilde{\theta}_{12}) - \re{(u)} \sin(2 \tilde{\theta}_{12})\right] 
%- \frac{\lambda^2}{2}\,\im{(u)} \sin(2 \tilde{\theta}_{12}) + \nonumber \\ &- \frac{\lambda^3}{4 \sqrt{2}}
%\left\{-A \eta \sin(2 \tilde{\theta}_{12})
%+ 2 A \eta \re{(u)}\left[-\cos(2 \tilde{\theta}_{12}) + \cos(\tilde{\theta}_{12}) \re{(u)} \sin(\tilde{\theta}_{12})\right] + 
%\nonumber \right. \\ \left.
%&+& \im{(u)} (-((1 + 2 A \rho) \cos(2 \tilde{\theta}_{12}) + 
%  3 A \eta \im{(u)} \sin(2 \tilde{\theta}_{12})\right\}
%\label{jarls}
%\end{eqnarray}
\begin{align}
&J_{\text CP}=\frac{\lambda}{4} \,\sigma\,
\im{(u)} \, \sin(2 \tilde{\theta}_{12})
+ \frac{\lambda^2}{2\sqrt{2}}\,\im{(u)} \cos(2 \tilde{\theta}_{12}) + \nonumber \\
& -\frac{\lambda^3}{8}\sigma\sin(2 \tilde{\theta}_{12})
\left[\sqrt{2} A \eta + 2 \im(u) \left(2 + |u|^2 + \sqrt{2} \re(u)\right) \right]\,.
\label{jarls}
\end{align}
Some comments are in order:
\begin{itemize}
\item in the limit of exact TBM, BM and GR, the invariant $J$ reduces to:
\begin{eqnarray}
J^{\text{TBM}}_{\text{CP}}=-\frac{A \eta \lambda^3 \sigma}{6}\,, \qquad J^{\text{BM}}_{\text{CP}}=-\frac{A \eta \lambda^3\sigma}{4 \sqrt{2}}\,, \qquad 
\qquad J^{\text{GR}}_{\text{CP}}=-\frac{A \eta \lambda^3\sigma}{2 \sqrt{10}}\,,
\end{eqnarray}
which all lead to a suppressed CP violation in the lepton sector, in agreement with Tab.(\ref{zeroth}) for an appropriate choice of $\sigma$;
%From a numerical point of view, for both ordering of the neutrino masses, $J^{TBM,BM,GR} \sim 0.02\, J^{exp}$, see Tab.({\ref{num}}), providing that $\sigma = +1$ for all patterns. However, we have to remark that the large corrections needed to reconcile $J$ with its experimental value can change the sign of $\sigma$;
\item retaining terms proportional to $\re(u)$ (and setting $\im(u)=0$) does not cure the previous problem since they appear only to ${\cal O}(\lambda^3)$;
\item to reconcile our prediction with the experimental value, we need to allow a deviation from exact TBM, BM and GR forms provided by $\im(u)$. The ${\cal O}(\lambda)$ degeneracy between $\sigma$ and $\im(u)$ will allow the latter to assume both positive and negative values.
\end{itemize}
\label{num}

\begin{figure}[b!]
  \begin{center}
    \includegraphics[scale=0.35]{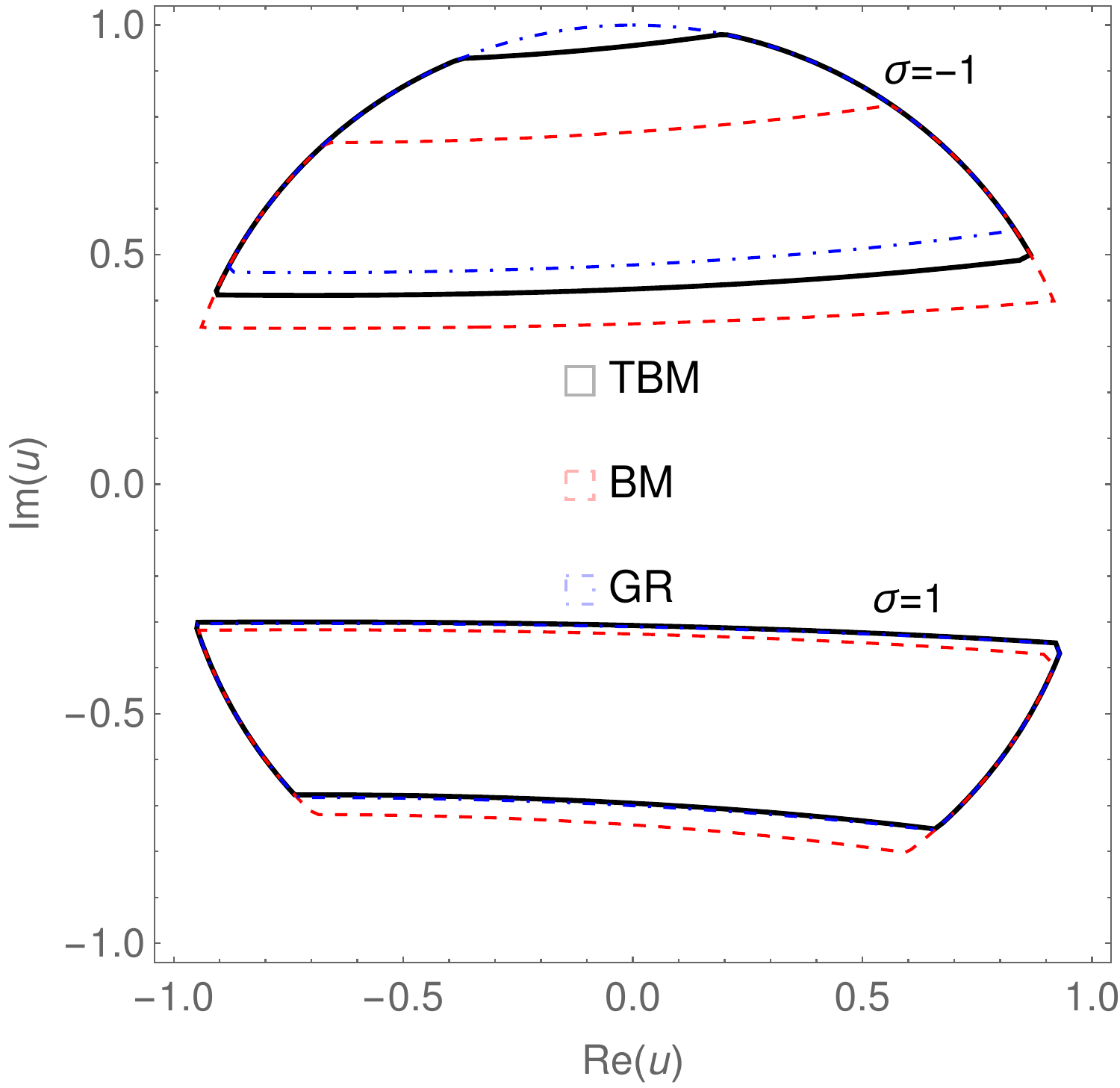}
    \caption{\small{\it Ensemble of $u$ values which make the modified versions of TBM (black solid line), BM (red dashed line) and GR (blue dot-dashed line) compatible with $J^{bf}_{\text{CP}}$ at 1$\sigma$. The upper (lower) plots show the solutions obtained with $\sigma=-1$ ($\sigma=+1$).}}
    \label{Dmaps}
  \end{center}
  \end{figure}

To find the set of values of $\re(u),\im(u)$ that allows to reproduce  the best fit point of $J_{\text{CP}}$ ($J^{bf}_{\text{CP}}$), in Fig.(\ref{Dmaps}) we plot the ensemble of $u$ values which makes the modified versions of TBM (black solid line), BM (red dashed line) and GR (blue dot-dashed line) compatible with $J^{\text{bf}}_{\text{CP}}$ at 1$\sigma$, subject to the constraint $|u|<1$. Given the similarities in the analytical structure of TBM, BM and GR, we see that overlapping complex $u$-region are covered; in addition, as commented above, we also expect a less relevant dependence on $\re(u)$ compared to $\im(u)$. 
The main conclusion is that $|\im(u)|\gtrsim 0.3$ (and  almost any $\re(u)$ in the $[-1,1]$ range) is enough to get the correct amount of leptonic CP violation, for any choice of the starting matrix. The mild $\re(u),\im(u)$ correlation is mainly dictated by the constraint $|u|<1$.
%\subsection{mixing angles}  

Now we go for the expressions of the mixing angles. 
%From a more general point of view, we have to verify that the allowed region in the $(\re(u),\im(u))$-plane does not destroy the prediction of the mixing angles. 
For the reactor angle we get a formula which is independent on the $\tilde{\theta}_{12}$ parameter (and thus on the sign of $\sigma$) up to $ {\cal O}(\lambda^3)$ terms (notice that ${\cal O}(\lambda^2)$ terms vanish):
\begin{eqnarray}
\label{reactor}
\sin(\theta_{13})&=&    \sqrt{1/2 + |u|^2 + \sqrt{2} \re(u)}\, \lambda+ \nonumber \\
&&\frac{\left[2 \sqrt{2} A \rho - 4 A \eta \im{u} + (-2 + 4 A \rho) \re{u} - 
  |u|^2 (3 \sqrt{2} + 2 \re(u))\right]}{4\sqrt{1 + 2|u|^2 + 2\sqrt{2} \re(u)}}\,\lambda^3\,.
\end{eqnarray}
In the limit of exact TBM, BM and GR mixing ($u=0$), we recover the well-known relation $\sin(\theta_{13}) = \frac{\lambda}{\sqrt{2}}+ {\cal O}(\lambda^3)$\,,
which is still a good approximation, see also Tab.(\ref{zeroth}). Moreover, eq.(\ref{reactor})  shows that, barring accidental cancellations, negative $\re(u)$ values are needed to compensate for positive shifts driven by $|u|$ (unless $\im(u)$ is also small, in that case small positive values of $\re(u)$ are also allowed).
%, otherwise the good relation in eq.(\ref{SMK}) would be badly broken.

Not too much must be said for the atmospheric angle; up to  ${\cal O}(\lambda^3)$ we get:
\begin{eqnarray}
\label{atm}
\tan(\theta_{23})&=&   1  + \frac{\lambda^2}{2} \left[ -1 + 4 A - 2 \sqrt{2}\re(u) \right] \,.
\end{eqnarray}
The most interesting feature is the absence of any dependence on $\im(u)$; thus, the small deviations from maximality are governed, beside the Cabibbo angle, by 
$\re(u)$ only. We also have to mention that the current best fit point is away from maximal mixing at the level of 3$\sigma$, see Tab.(\ref{nufit_table}). Thus, relatively large positive $\re(u)$ are needed to shift $\tan(\theta_{23})$ towards its 1$\sigma$ preferred value which lies around $\tan(\theta_{23})^{bf}\sim 0.9$.
As in the previous case, no dependence on $\tilde{\theta}_{12}$ appears so that exact TBM, BM and GR hypothesis give the same expression in eq.(\ref{atm})  with $\re(u) =0$.
%\footnote{Notice that the expression for $\theta_{13}$ quoted in Tab.(\ref{zeroth}) is just a leading order in $\lambda$.}:
%\begin{eqnarray}
%\label{atm2}
%\tan(\theta_{23})&=&   1 +\frac{\lambda^2}{2} \left[ -1 + 4 A\right]   + {\cal O}(\lambda^4)
%\nonumber \,.
%\end{eqnarray}

Finally, for the solar angle we get:
\begin{eqnarray}
\label{sol}
\tan(\theta_{12})&=&  \tan(\tilde{\theta}_{12}) + \frac{\lambda}{\sqrt{2} \tilde{c}^2_{12}}\sigma +
 \frac{\lambda^2}{2 \tilde{c}^3_{12}} \tilde{s}_{12} + \nonumber \\
 &+& \frac{\lambda^3}{4\tilde{c}^4_{12}} \sigma\,
 \left[\sqrt{2} (1 - 2 A \tilde{c}_{12}^2 \rho) + \tilde{c}_{12}^2 (\sqrt{2} |u|^2 + 2 \re(u)) \right]\,.  
\end{eqnarray}
The most considerable feature is that the corrections implied by $U_{13}$ of eq.(\ref{notpmns2}) are too small to be significant; thus, the expressions of $\theta_{12}$ are very similar to those quoted in Tab.(\ref{zeroth}). In addition, 
%as expected, there is a strong dependence on $\tilde{\theta}_{12}$, so the predictions from TBM, BM and GR 
%limits displayed in Tab.(\ref{zeroth}) are different; in particular, for the function $\tan(\theta_{12})$ and $\lambda=0$, we recover:
%\begin{equation}
%\label{ccaler}
%\tan(\theta_{12})^{\lambda = 0} =
%\begin{cases}
%\frac{1}{\sqrt{2}} \qquad {\rm for \;TBM}\\ 1  \; \;\qquad {\rm for \;BM} \\
%2 \sqrt{5}/(5 + \sqrt{5})\sim 0.62\qquad {\rm for \;GR} 
%\end{cases}
%\end{equation}
%to be compared with the experimental best fit value $\tan(\theta_{12})^{bf}\sim 0.66$. Thus, ${\cal O}(\lambda)$ corrections 
%are crucial to reconcile the ansatz predictions with experiments. Secondly, 
once we specify the values of $\tilde{\theta}_{12}$ for the relevant patterns, there are no free parameters up to ${\cal O}(\lambda^2)$; we can then derive the following sum-rules among physical angles (that, for the sake of simplicity, we report here up to first order in $\sin(\theta_{13})$):
\begin{equation}
\label{ccaler2}
\tan(\theta_{12}) =
\begin{cases}
\frac{1}{\sqrt{2}} +   3 \sigma\,\sin(\theta_{13})/2
\qquad {\rm for \;TBM}\\ 1 + 2 \sigma\sin(\theta_{13}) \; \;\qquad {\rm for \;BM} \\
2 \sqrt{5}/(5 + \sqrt{5}) + 10\sigma \sin(\theta_{13})/(5 + \sqrt{5})\qquad {\rm for \;GR} \,.
\end{cases}
\end{equation}
The only possibility to (marginally) reconcile the previous sum rules with the experimental value happens for BM mixing with $\sigma=-1$, which shows a deviation from $\tan(\theta_{12})^{bf}$ at around $\sim 3\%$ (compare with Fig.(\ref{th12bar})); 
for the other mixing patterns, this difference amounts to values as large as $\sim$20\% for GR and  $\sim$30\% for TBM. 
%The reason is that a correction of ${\cal O}(\sin(\theta_{13}))$ is too large for those patterns whose solar angle is already very close to the measured value. 
To better quantify the (dis-)agreements of the obtained $U_{\text{PMNS}}$ with the experimental data after including the corrections in eq.(\ref{notpmns2}), we perform a simple $\chi^2$ test, with the function:
\begin{equation}
\resizebox{.9 \textwidth}{!}{$\chi^2=\frac{\left[J_{\text{CP}}-J_{\text{CP}}^{bf}\right]^2}{\sigma_{J_{\text{CP}}}^2} +  \frac{\left[\sin(\theta_{13})-\sin(\theta_{13})^{bf}\right]^2}{\sigma_{\sin(\theta_{13})}^2} +  \frac{\left[\tan(\theta_{23})-\tan(\theta_{23})^{bf}\right]^2}{\sigma_{\tan(\theta_{23})}^2} +\frac{\left[\tan(\theta_{12})-\tan(\theta_{12})^{bf}\right]^2}{\sigma_{\tan(\theta_{12})}^2}$ }\label{chi2}
\end{equation}
%where, with obvious notation, the superscript $"bf"$ refer to the best fit values quoted in Tab.(\ref{nufit_table}).
For all patterns, the minimum of the $\chi^2$ is very large, in the range $(10-10^3)$ and it is dominated by the $\tan(\theta_{12})$ term; in fact,  
if we exclude $\theta_{12}$ from the $\chi^2$ function, the fit improves considerably for all patterns, with $\chi^2_{min}\sim {\cal O}(20)$ (the best performance being obtained by  BM mixing with $\sigma=-1$). The problem  related to the deviation from maximality of $\theta_{23}$ is, instead, less relevant because of a larger relative 1$\sigma$ error compared to $\theta_{12}$. 
Finally, the corrections analyzed here help in improving the values of $\Delta$, $\Delta=(2708, 2492,2573)$ for TBM, BM and GR mixings, respectively.
%\begin{figure}[h!]
%  \begin{center}
%    \includegraphics[scale=0.5]{figures/bm_fake.pdf}
%    \caption{\small{\it 99.5\% CL and 99.9\% fit of the modified BM mixing to the neutrino mixing angles but $\theta_{13}$.}}
%    \label{bmfit}
%  \end{center}
%  \end{figure}
Obviously, assuming for the variable $u$ a smaller value, that is shifting $u \to \lambda^N \,u$,  does not solve the problem for any integer $N$.

\subsection{Perturbation on the (23)- sector}
\begin{figure}[h!]
  \begin{center}
    \includegraphics[scale=0.35]{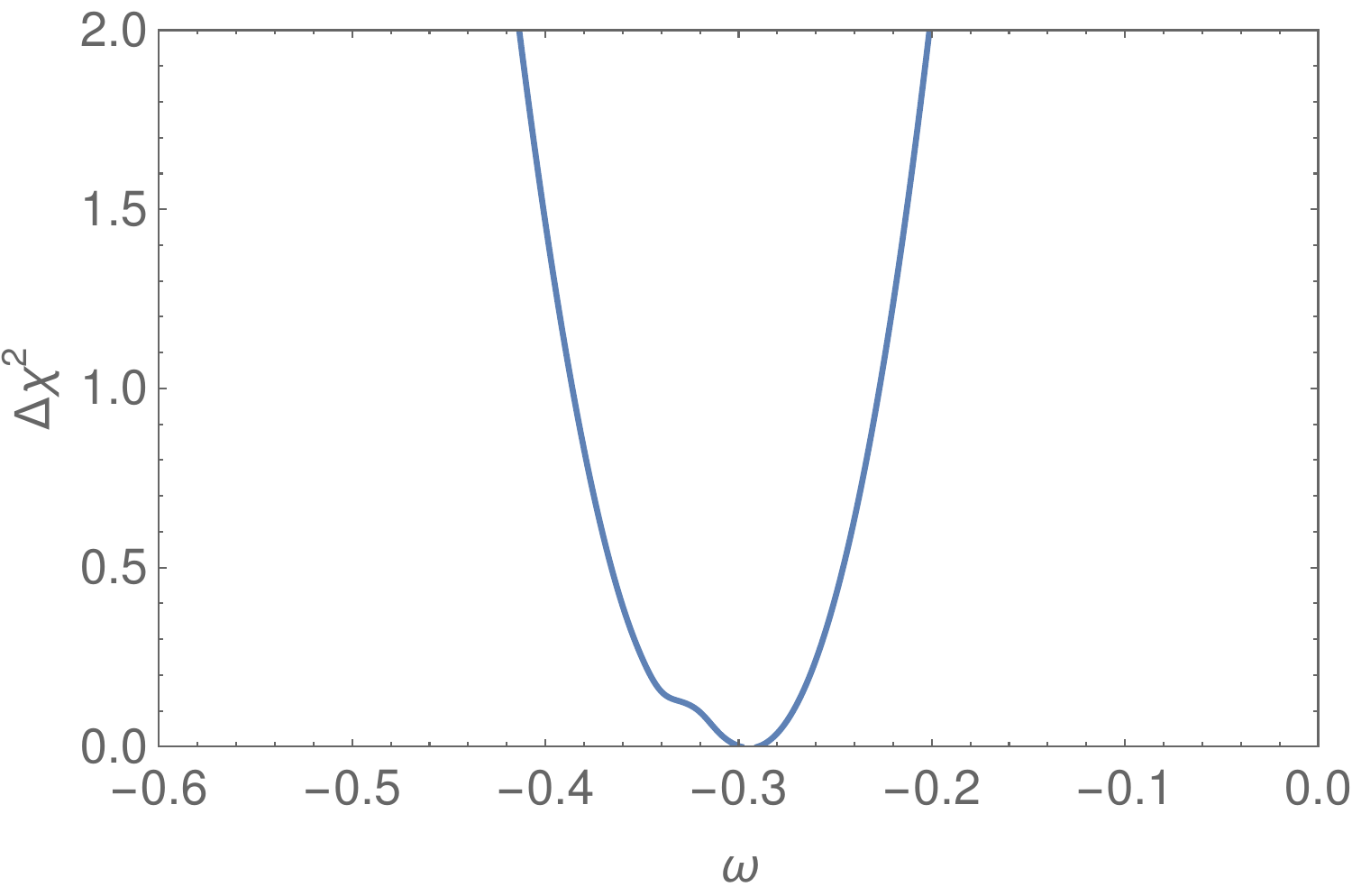}
    \caption{\small{\it$\Delta \chi^2 = \chi^2-\chi^2_{min}$ behaviour as a function of $\omega$ for the modified BM mixing. In the fit procedure, we have marginalized over the 
    $(\re(u),\im(u)))$ pair.}}
    \label{bmfit2}
  \end{center}
  \end{figure}

One possibility to alleviate the problem in the (23)-sector is to slightly modify $U_{23}$ of eq.(\ref{defT}) by inserting a new real parameter $\omega$ according to\footnote{Since the complex  variable $u$ was already enough to guarantee the correct amount of leptonic CP violation, we prefer to reduce the number of free parameters choosing a real correction $\omega$.}:
\begin{equation}
\label{u23}
U_{23} =
\begin{pmatrix}
1 &0  & 0\\
0 & \frac{1}{\sqrt{2}}-\lambda \omega - \sqrt{2} \lambda^2 \omega^2 - 
 2 \lambda^3 \omega^3
 & \frac{1}{\sqrt{2}} +\omega \lambda\\
0 &
-\frac{1}{\sqrt{2}} -\omega \lambda& \frac{1}{\sqrt{2}} -\lambda \omega - \sqrt{2} \lambda^2 \omega^2 - 
 2 \lambda^3 \omega^3
\end{pmatrix}\,.
\end{equation}
Notice that, to maintain the unitarity of $U_{23}$, we displayed up to ${\cal O}(\lambda^3)$ terms.
We repeat the same calculations as before and indicate with a {\it prime} the new expressions of the mixing parameters while leaving {\it unprimed} the results of the previous section. The relevant corrections driven by $\omega$ are as follows:
\begin{eqnarray}
 J_{\text{CP}}^\prime &=& J_{\text{CP}} -\frac{\lambda^3}{2} \,\im(u)\,\left[\omega \cos\left(2 \tilde{\theta}_{12}\right) + 2 \sigma \omega^2 \sin\left(2 \tilde{\theta}_{12} \right)\right]\nonumber \\
 \sin^\prime(\theta_{13})&=& \sin(\theta_{13}) + \frac{\lambda^2 \omega}{\sqrt{2}} \,\frac{\left[\sqrt{2} + 2 \re(u)\right] }
 {\sqrt{1 + 2 |u|^2 + 2 \sqrt{2} \re(u)}} \nonumber \\ \label{secit}
 \tan^\prime(\theta_{23}) &=& \tan(\theta_{23}) + 2 \sqrt{2} \lambda \omega\\
 \tan^\prime(\theta_{12}) &=& \tan(\theta_{12}) - \frac{\lambda^2 \omega \sigma}{\cos^2\left(\tilde{\theta}_{12}\right)}\nonumber \,.
\end{eqnarray}
We see that $J_{\text{CP}}, \theta_{12}$ and $\theta_{13}$ acquire small ${\cal O}(\lambda^{2-3})$ corrections that do not improve the fit compared to the previous section. For the atmospheric angle, instead, an ${\cal O}(\lambda)$ is relevant, especially for negative values of $\omega$ as, starting from maximality, we need a negative correction to jump into the experimental value{\footnote{The three main neutrino global fits \cite{Esteban:2020cvm,deSalas:2020pgw,Capozzi:2021fjo} do not agree on the preferred $\theta_{23}$ octant, even though the $3\sigma$ ranges are all compatible. In our analysis, an higher octant value for $\theta_{23}$ can be easily obtained with a positive $\omega$ value.}}. Notice that this is true for any value of $\sigma$.
However, even though the atmospheric angle turns out to be in the correct range, the fits to the expressions in eq.(\ref{secit}) are only slightly improved but still remain  $\gtrsim {\cal O}(100)$ because of the poor foreseen solar angle; as before, only the modified BM mixing case presents a good minimum of the  $\chi^2$ at $\chi^2_{min}=3.47$. For the sake of illustration, the behaviour of the $\Delta \chi^2 = \chi^2-\chi^2_{min}$  as a function of $\omega$ is presented in Fig.(\ref{bmfit2}). For every $\omega$, we have marginalized over $\re(u)$ and $\im(u)$ in the fit.
\subsection{The full glory: perturbation on the (12)- sector}
\begin{figure}[H]
    \centering
    \includegraphics[width=.323\textwidth]{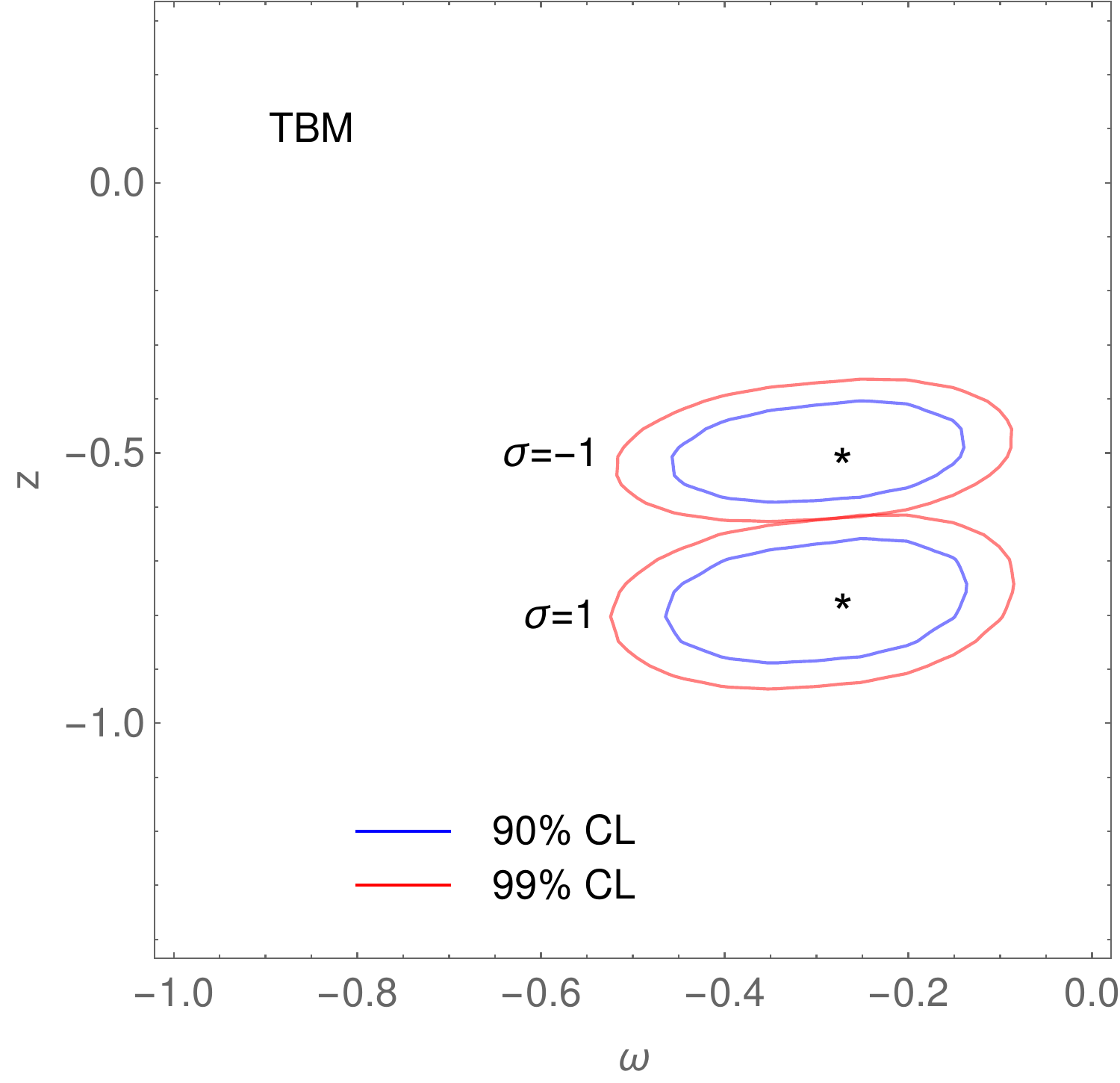}
    \includegraphics[width=.323\textwidth]{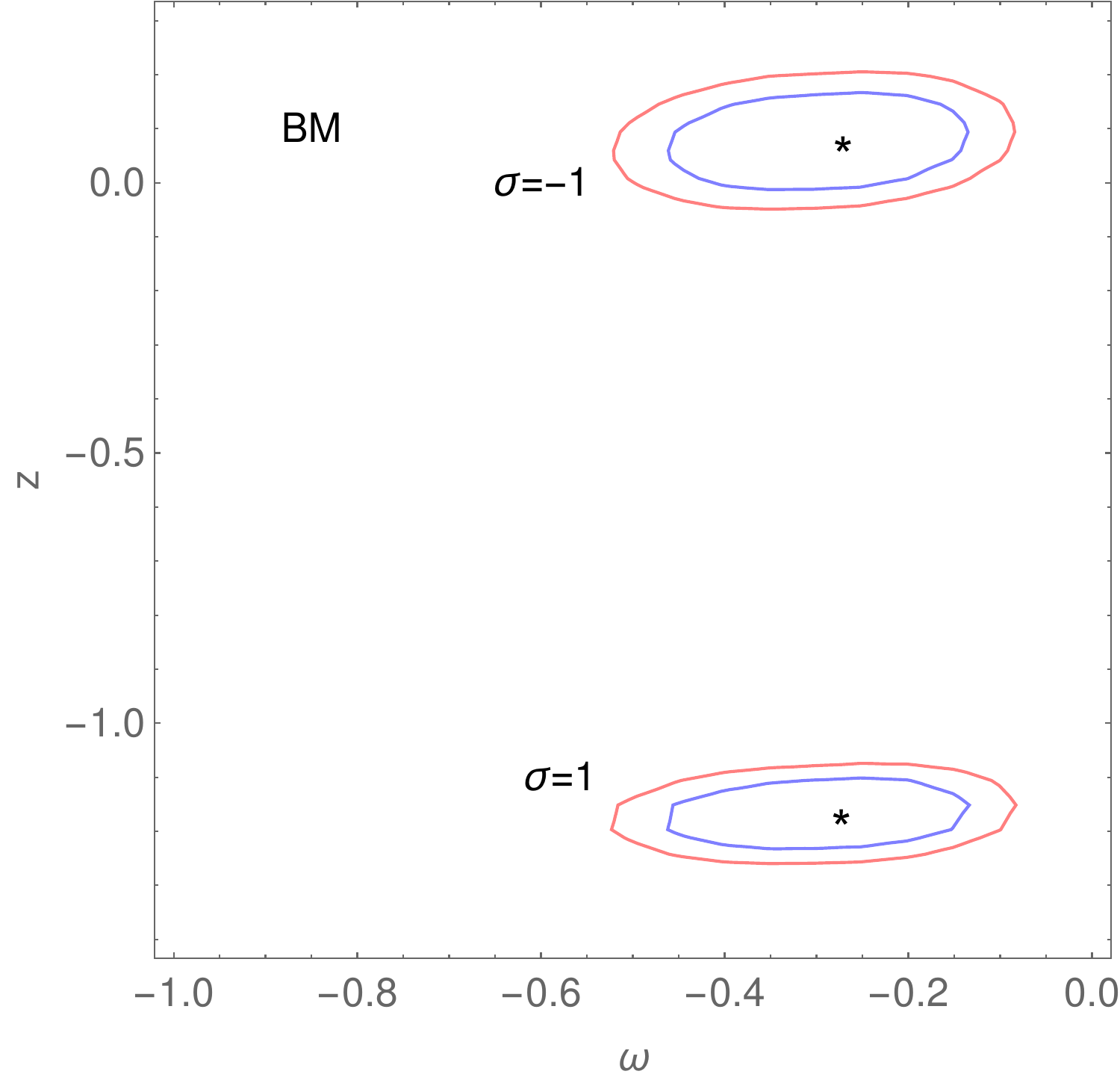}
    \includegraphics[width=.323\textwidth]{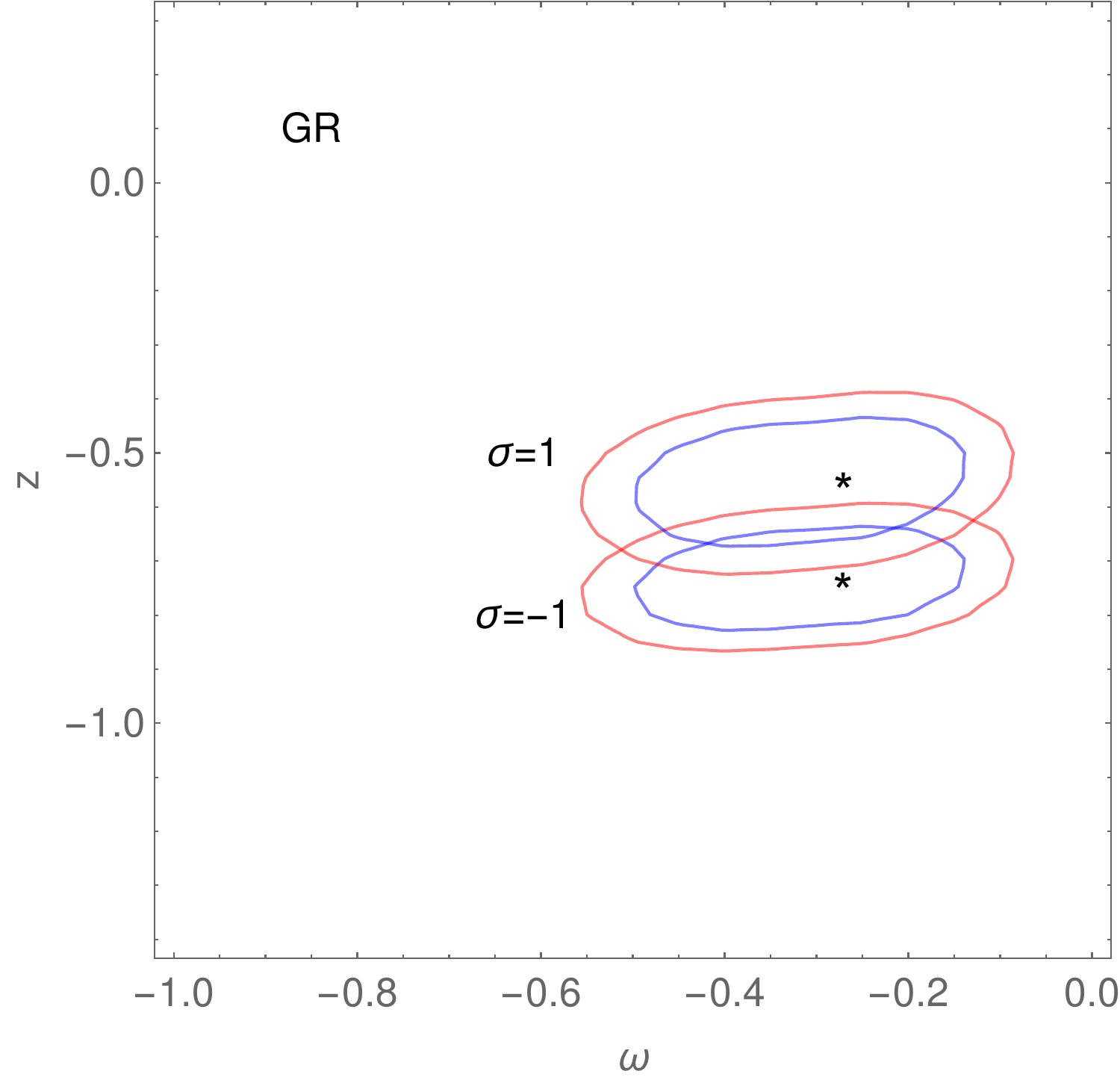}
    \caption{\it The 90\% and 99\% $\chi^2$ confidence levels in the $(\omega,z)$-plane for TBM (left panel), BM (middle panel) and GR (right panel). In each plots, we have reported  both $\pm 1$ possibilities for $\sigma$ and marginalized over the 
    $(\re(u),\im(u)))$ pair. 
    }
    \label{omegaZ}
\end{figure}

The results of the previous sections have shown that the predictions for $J_{\text{CP}}$ and $\sin(\theta_{13})$ are good for all mixing once the $u$-corrections are included. The $\omega$ corrections are needed to reconcile the $\theta_{23}$ deviations from maximal mixing (common to all patterns) while the solar angle remains sensitively away from its experimental value for TBM and GR mixing but sufficiently close to it for BM. Thus, in order to complete our program to match the data of Tab.(\ref{nufit_table}),  we need to add a (real) correction of ${\cal O}(\lambda)$ to the (12)-sector, that we dub with $z$. We parameterize it in the following way:
\begin{equation}
\label{new12}
U_{12} =
\begin{pmatrix}
K&
\tilde{s}_{12} \sigma+z \lambda   & 0\\
-\tilde{s}_{12} \sigma-z \lambda &  K & 0\\
0 &
0&1
\end{pmatrix}\,,
\end{equation}
where $K=\tilde{c}_{12} - \tilde{s}_{12} z  \sigma/\tilde{c}_{12} \lambda-z^2/(2 \tilde{c}_{12}^3) \lambda^2-  \tilde{s}_{12}  z^3  \sigma/(2 \tilde{c}_{12}^5) \lambda^3$.
The expression of the mixing parameters are modified accordingly; in particular, 
$\theta_{13}$ and $\theta_{23}$ are unaffected by $z$, so their expressions of eq.(\ref{secit}) are valid even in this case. The Jarlskog invariant gets an ${\cal O}(\lambda^2)$ correction of the form:
\begin{equation}
J_{\text{CP}}^{\prime\prime} = J_{\text{CP}}^\prime +     \frac{\lambda^2}{2}\,
\frac{\cos(2 \tilde{\theta}_{12}) \im(u)}{\cos(\tilde{\theta}_{12})}\,z \,.
\end{equation}
By construction, the most interesting case is related to $\theta_{12}$; here, corrections of ${\cal O}(\lambda)$ driven by $z$ compete with that shown in eq.(\ref{sol}):
\begin{eqnarray}
 \tan^{\prime\prime}(\theta_{12}) &=& \tan(\theta_{12})^\prime + \lambda\,\frac{ \sigma\,z}{\cos^3\left(\tilde{\theta}_{12}\right)}\label{final12} \,.
\end{eqnarray}
Thus, we expect that   a cancellation among the $\lambda$ coefficients could bring the TBM and GR mixing in agreement with the data (for any $\sigma$) while for BM the contribution from $z$ (and $\sigma=-1$) must be small in order not to destroy the agreement found above; conversely, we expect that  $\sigma=1$ will be acceptable for non-vanishing $z$ corrections.
To check whether this is the case, we minimized the $\chi^2$ function of eq.(\ref{chi2}) over the four independent parameters $\re(u),\im(u), \omega$ and $z$ and reported their best fit values  in Tab.(\ref{fit_table2}).
\begin{table}[h]
\centering
\renewcommand{\arraystretch}{1.2}
\begin{tabular}{l c c c c} 
\toprule
Pattern & $\re(u)$ & $\im(u)$ & $\omega$ & $z$ \\ 
\midrule
TBM & -0.27(-0.27) & 0.57(-0.55) & -0.27(-0.27)&  -0.50(-0.77) \\
BM &-0.27(-0.29) &0.57(-0.56) & -0.27(-0.27)& 0.08(-1.17)\\
GR & -0.27(-0.27) & 0.57(-0.54) & -0.27(-0.27)&  -0.73(-0.55) \\
\bottomrule
\end{tabular}
\caption{\small{\it{Values of the parameters  $\re(u),\im(u), \omega$ and $z$ that minimize the 
$\chi^2$ function of eq.(\ref{chi2}), computed for $\sigma=-1$ and, in parenthesis, for $\sigma=1$. For all patterns, $\chi^2_{min}\sim 0$.}
}}
\label{fit_table2}
\end{table}
 For all patterns, the minimum of the $\chi^2$ is very close to zero, so we did not report it on the table. As expected, the magnitude and signs of the needed $z's$ reflects our considerations below eq.(\ref{final12}). In addition, the very similar values for $\re(u)$ and $\im(u)$ can be understood from Fig.(\ref{Dmaps}), where the acceptable regions for such parameters are almost equivalent for each pattern. Finally, compared to the previous section, the value of $\omega$ is compatible with the  BM  case previously analyzed and, as expected, tends to assume a very similar strength for all other patterns and signs of $\sigma$ (${\cal O}(\lambda)$ corrections are universal).
The 90\% and 99\% confidence levels of the $\chi^2$ function in the $(\omega,z)$-plane for TBM (left panel), BM (middle panel) and GR (right panel) are reported in Fig.(\ref{omegaZ}); in each plots we included both $\pm 1$ possibilities for $\sigma$ and marginalized over the 
$(\re(u),\im(u)))$ pair.
%\begin{figure}[H]
%    \centering
%    \includegraphics[width=.4\textwidth]{figures/TBM_full_plus1OmegaZ.pdf}
%    \includegraphics[width=.4\textwidth]{figures/TBM_full_minus1OmegaZ.pdf}\\
%    \includegraphics[width=.4\textwidth]{figures/BM_full_plus1OmegaZ.pdf}
%    \includegraphics[width=.4\textwidth]{figures/BM_full_minus1OmegaZ.pdf}\\
%    \includegraphics[width=.4\textwidth]{figures/GR_full_plus1OmegaZ.pdf}
%    \includegraphics[width=.4\textwidth]{figures/GR_full_minus1OmegaZ.pdf}
%    \caption{Fits in the $(\omega,z)$-plane.
%    }
%    \label{omegaZ}
%\end{figure}

\section{On the neutrino masses}
\label{neumasses}
\begin{figure}[H]
    \centering
    \includegraphics[width=.25\textwidth]{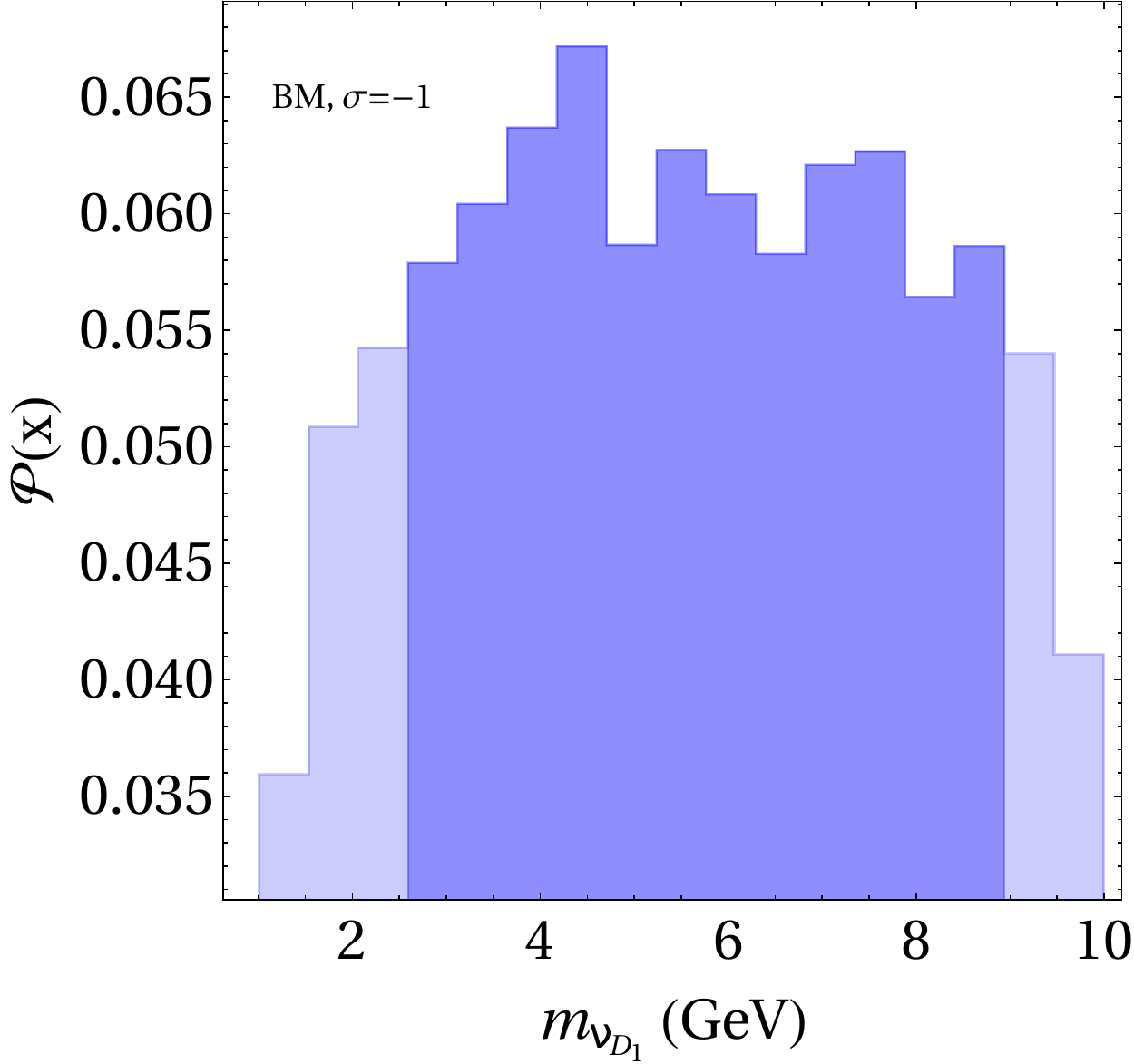}
    \includegraphics[width=.25\textwidth]{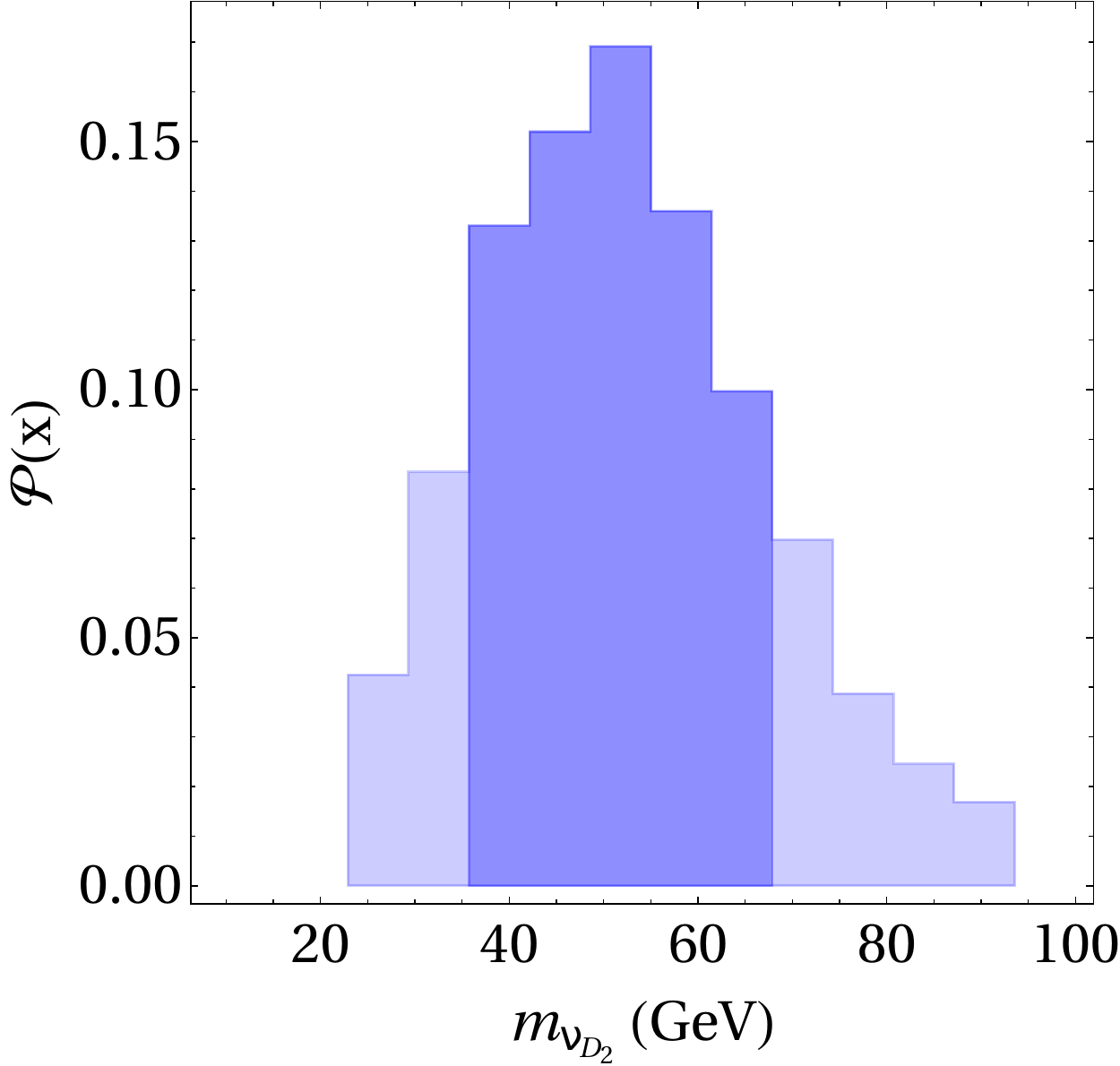}
    \includegraphics[width=.25\textwidth]{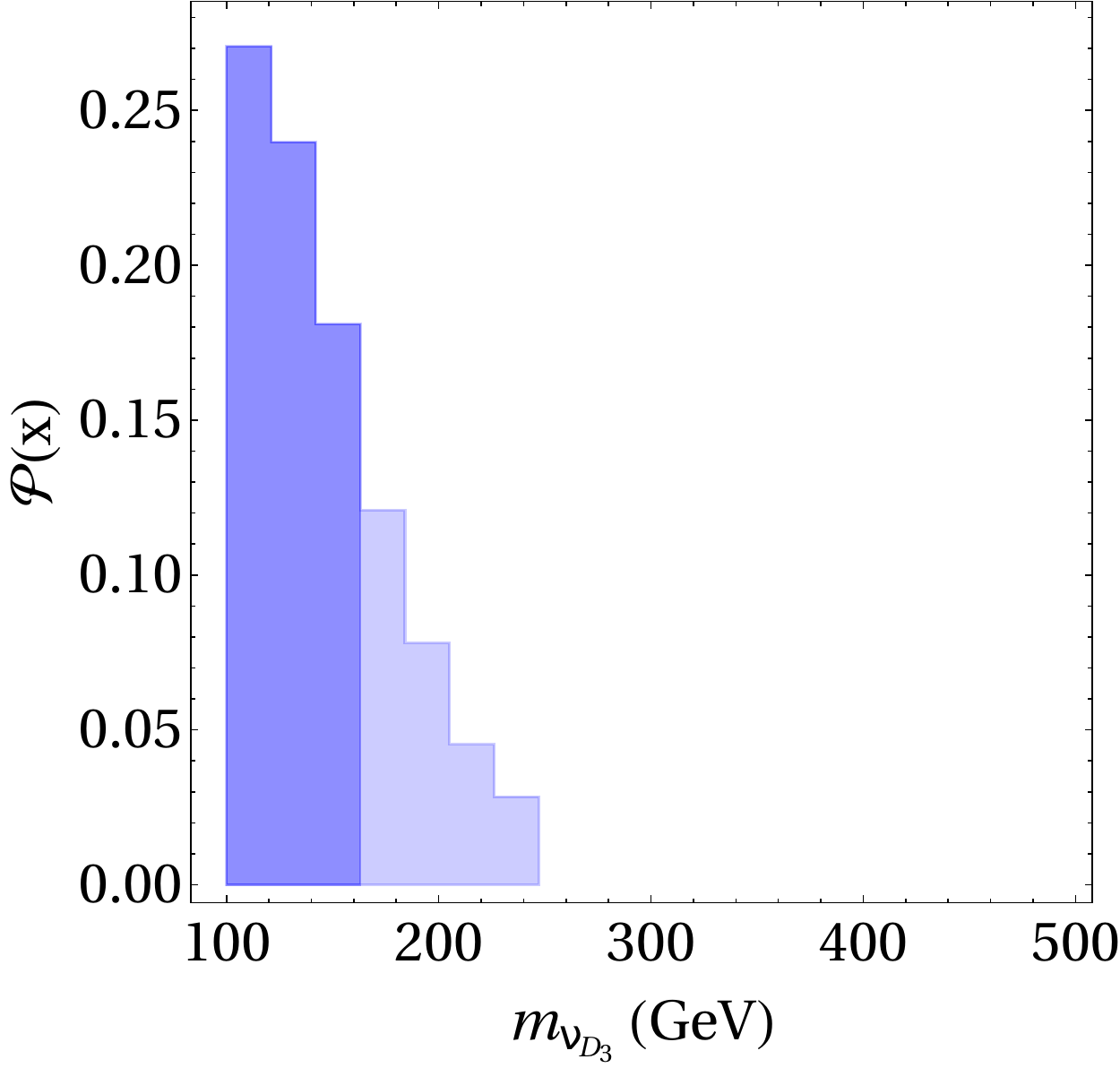}\\
    \includegraphics[width=.25\textwidth]{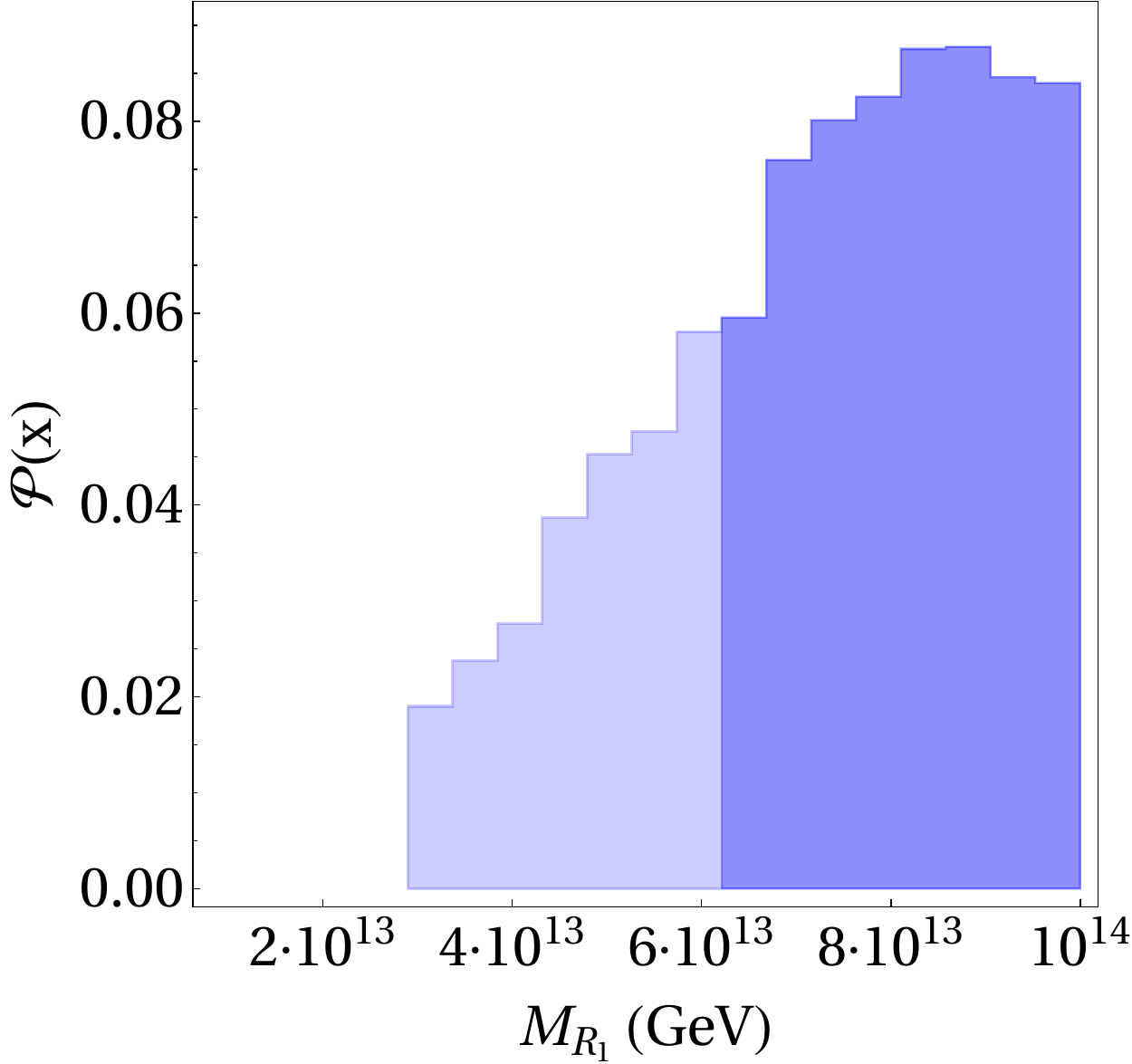}
    \includegraphics[width=.25\textwidth]{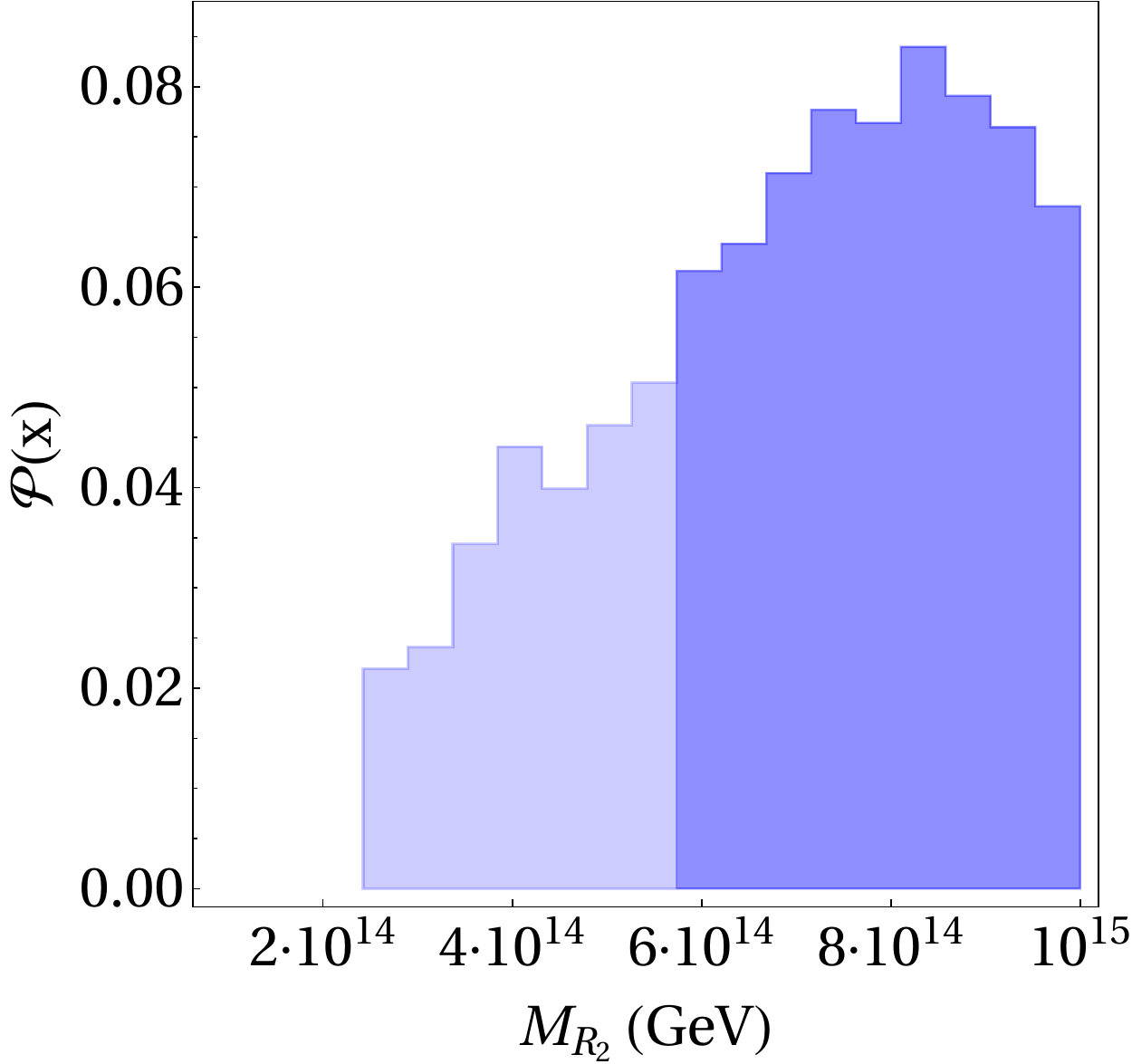}
    \includegraphics[width=.25\textwidth]{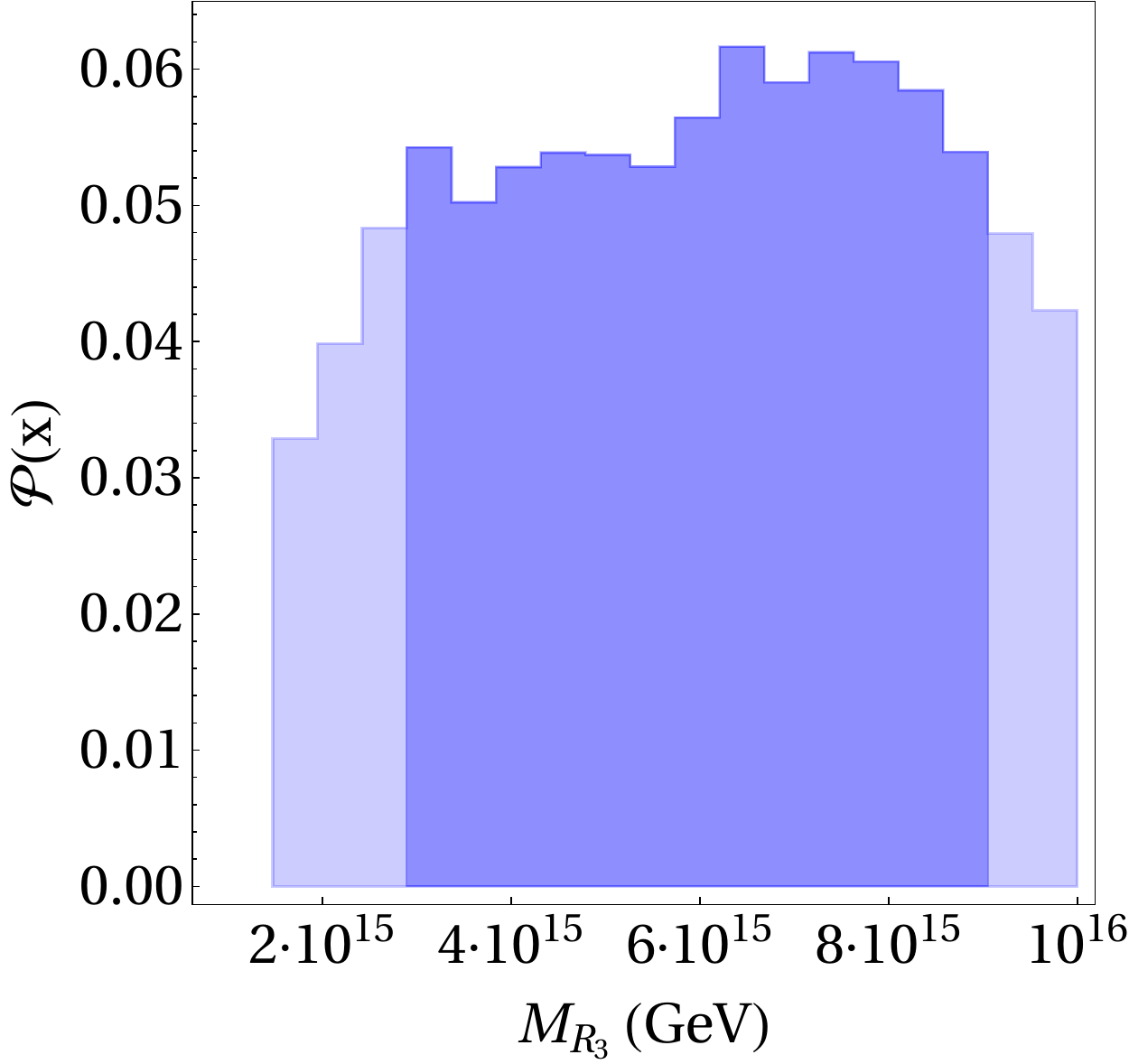}\\
    \includegraphics[width=.25\textwidth]{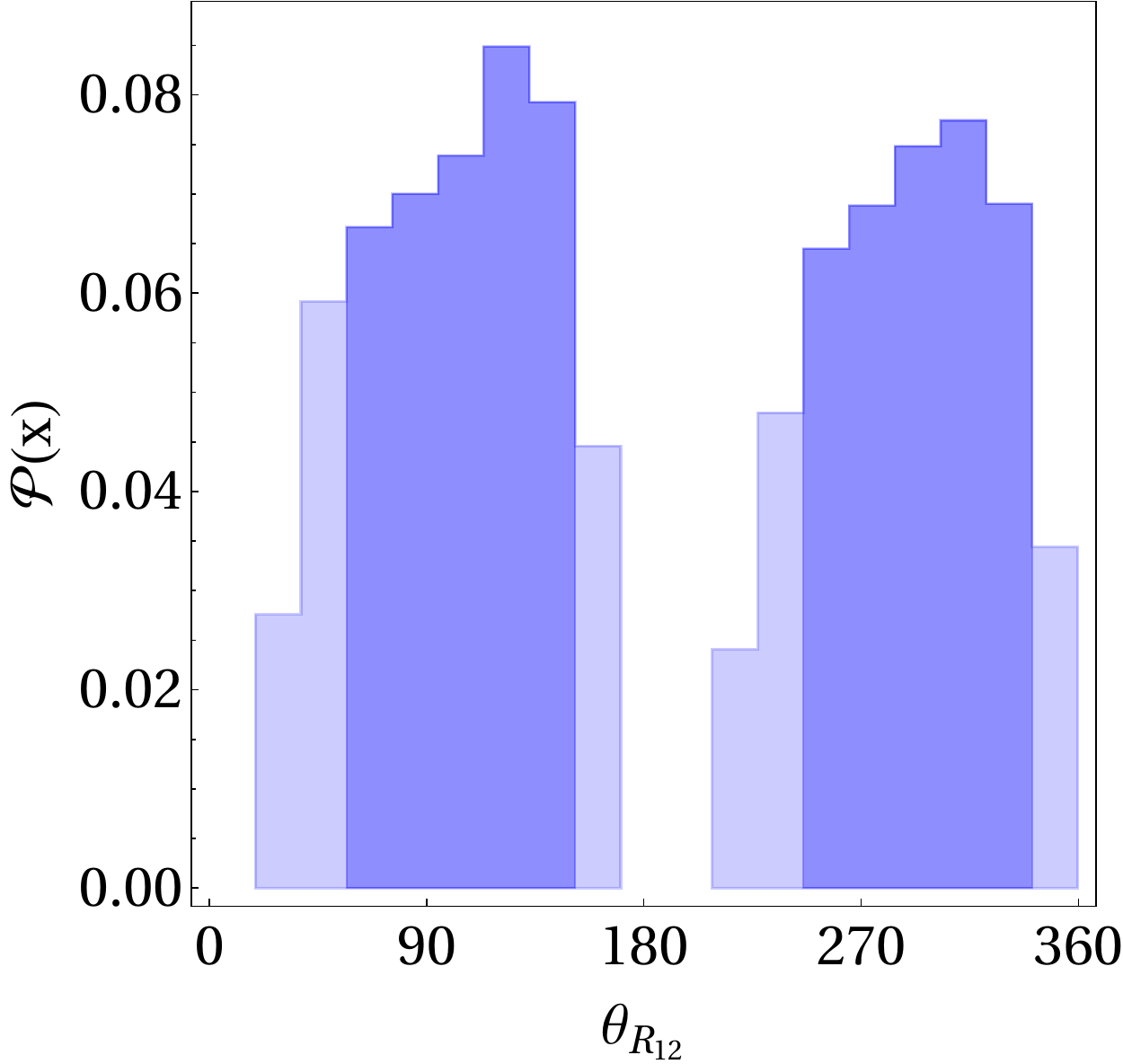}
    \includegraphics[width=.25\textwidth]{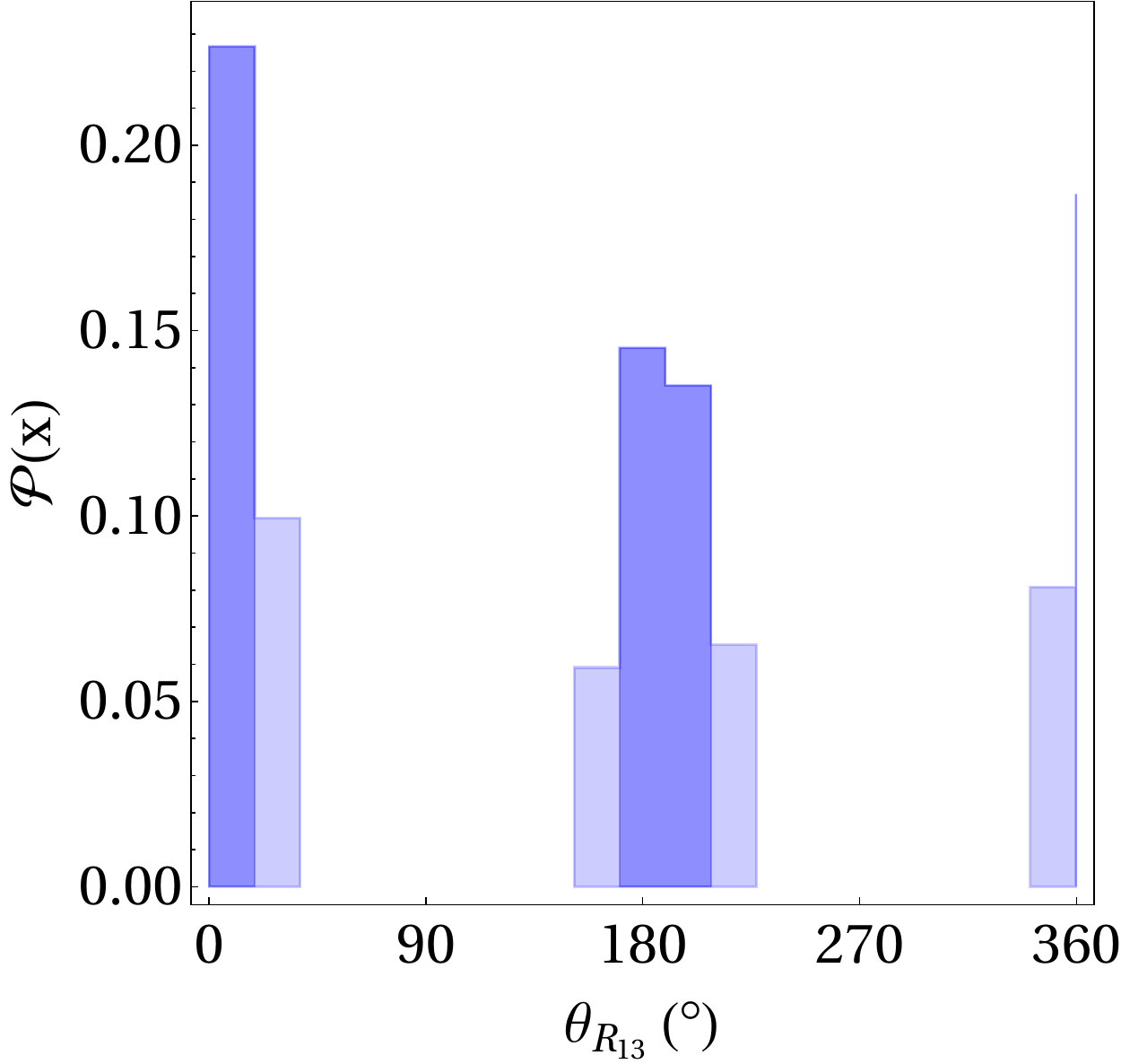}
    \includegraphics[width=.25\textwidth]{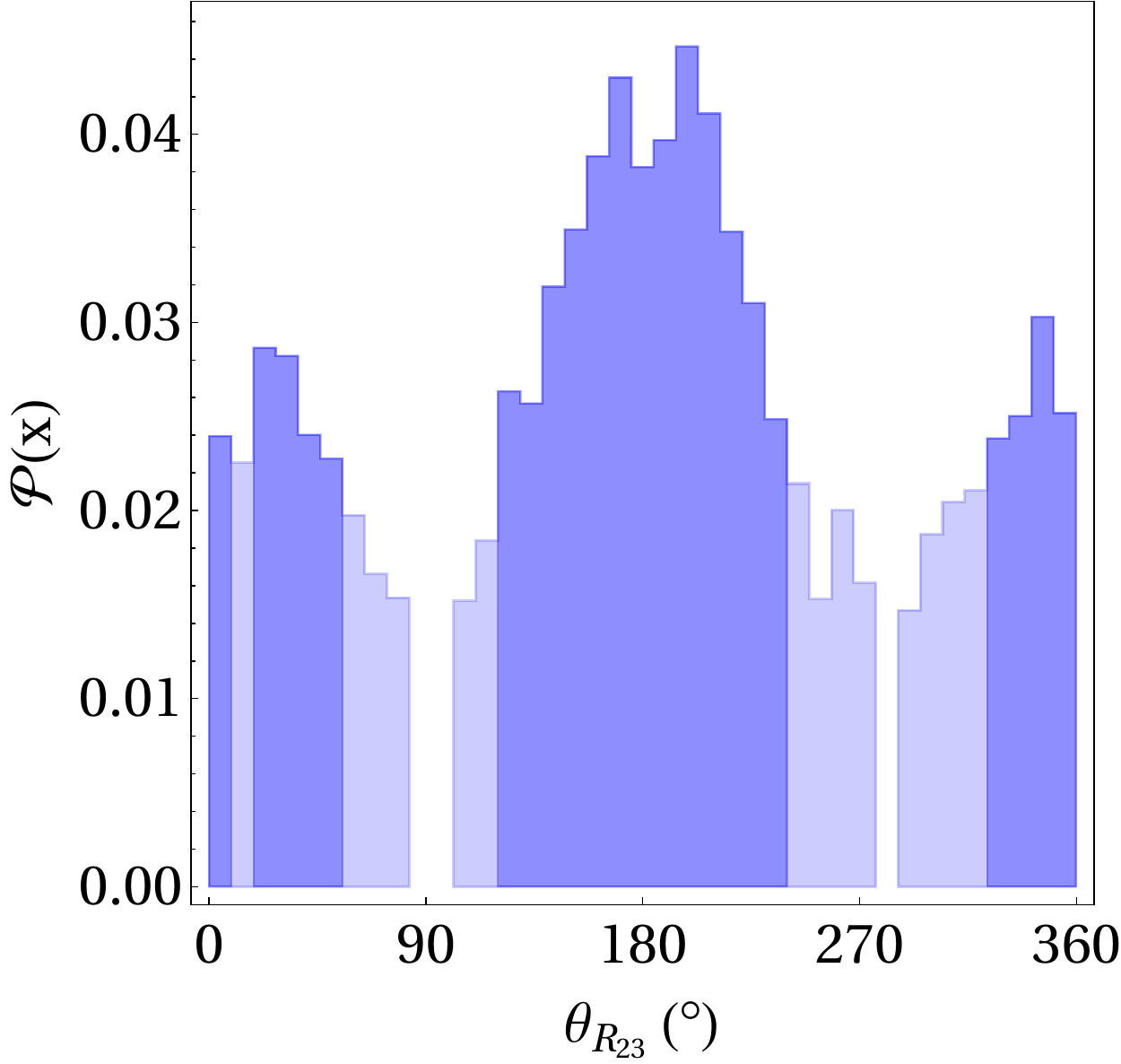}
    \caption{\it Posterior distributions for the elements of the vectors $\vec{m}_{\nu_{D}}$ (first line) $\vec{M}_{R}$ (middle line) and $\vec{\theta}_R$ (lower line) for the BM case, $\sigma=-1$. Darker and lighter blue refer to 68\% and 95\% credible intervals, respectively.
    }
    \label{posterior}
\end{figure}

The next step is to ensure that our procedure is able to reproduce the solar and atmospheric mass differences. Eq.(\ref{eq1}) offers the structure of the neutrino mixing matrix in terms of a right rotation $U_R$ (four real parameters), three right-handed neutrino masses and three Dirac neutrino masses, for a total of ten unknown parameters; of those, four have been used to constrain the matrix $T$ in eq.(\ref{m0}), and the remaining  six parameters are left to describe neutrino masses. To determine them, one can try to figure out the structure of the diagonal matrix $S$ by inverting eq.(\ref{m0}), so that:
\begin{eqnarray}
\label{smatrix}
S=T^\dagger  m^{diag}_{\nu_D} \,U_R^\dagger \,(M_R^{diag})^{-1}\, U_R^\star m^{diag}_{\nu_D} T^\star \,.
\end{eqnarray}
Notice that the matrix $S\equiv m_\nu^{diag}$ does not depend on the quark mixing. 
One possibility to determine the unknown parameters is to rephrase eq.(\ref{smatrix}) to the more useful form:
\begin{eqnarray}
\label{smatrix2}
S-T^\dagger m_0 T^\star=0\,.
\end{eqnarray}
Its left-hand side is a symmetric matrix made complex by the entries of $ m^{diag}_{\nu}$ and by the $T$ matrix, needed to successfully reproduce the leptonic CP violation. Thus, eq.(\ref{smatrix2}) is equivalent to 12 conditions, which have to be simultaneously valid. 
However, we can easily verify that the imaginary parts of the elements of $T$ are always smaller than the real part (at the level of 20\% or smaller) with a notable exception of the element (13), for which the imaginary part is either larger (in the only case when $T$ is the corrected BM mixing with $\sigma=-1$) or just half of the real part. With the aim of catching the relevant physics, not obfuscated by useless details (phases are of the uttermost importance for CP violation, not for neutrino masses), we prefer to deal with real $S$ and $T$ matrices; this allows us to reduce the number of constraints to six only\footnote{If, instead, we prefer to deal with complex matrices, thus phases must be added to $U_R$ that helps in making vanishing all imaginary parts of eq.(\ref{smatrix2}).}. Even in this case,  the large number of free parameters makes the expressions of neutrino masses quite cumbersome. Thus, we only give a numerical solution to eq.(\ref{smatrix2}). For the $S$ matrix we take the following expression, valid for the Normal Ordering (NO) case:
\begin{equation}
\label{new}
S=\diag(m_1,\sqrt{m_1^2+\Delta m^2_{\text sol}},\sqrt{m_1^2+\Delta m^2_{\text atm}})\,,
\end{equation}
where $m_1$ is the absolute neutrino mass scale that, for the sake of simplicity, we assume vanishing. We then construct the {\it adimensional} function:
\begin{eqnarray}
\label{ffunction}
F(\vec{m}_{\nu_{D}},\vec{M}_{R},\vec{\theta}_R)=\frac{\sum_{j<i=1}^3  \left[S_{ij}-(T^T m_0 T)_{ij}\right]^2}{\Delta m^2_{\text{sol}}} \,,
\end{eqnarray}
and look for minima as close as possible to zero. Here the vectors have the following entries, with obvious meaning:
\begin{eqnarray}
\vec{m}_{\nu_{D}} = \left(m_{\nu_{D_1}},m_{\nu_{D_2}},m_{\nu_{D_3}}\right) \qquad 
\vec{M}_{R}=\left(M_{R_1},M_{R_2},M_{R_3}\right)\qquad \vec{\theta}_R=\left(\theta_{R_{12}},\theta_{R_{13}},\theta_{R_{23}}\right)\,.
\label{vectors}    
\end{eqnarray}
We consider ourselves satisfied when 
$F(\vec{m}_{\nu_{D}},\vec{M}_{R},\vec{\theta}_R)<1$, meaning that all the differences between the corresponding matrix elements of $S$ and $T^T m_0 T$ are smaller than the 
smallest measured mass scale $\Delta m^2_{\text{sol}}$.
The minimization procedure has been carried out by means of the software MultiNest, which is based on nested sampling normally used for calculation of
the Bayesian evidence \cite{Feroz:2007kg,Feroz:2008xx,Feroz:2013hea}.
The choice of priors in this context is relevant. To prove that a solution to the system (\ref{smatrix2}) exists, we set:
\begin{alignat}{3}
1 \le m_{\nu_{D_1}}/{\text GeV}&<  10\,, \qquad  & 10\le m_{\nu_{D_2}}/{\text GeV}<  100\,, 
\qquad   100\le m_{\nu_{D_3}}/{\text GeV}&<  500\,,  \nonumber \\ \nonumber \\
10^{13}\le M_{R_1}/{\text GeV}&<  10^{14}\,, \qquad  &10^{14}\le M_{R_2}/{\text GeV}<  10^{15}\,, 
\qquad   10^{15}\le M_{R_3}/{\text GeV}&<  10^{16}\,, \\ \nonumber \\
\phantom{\leq m_{\nu_{D_1}}/{\text GeV}}&\phantom{<  10\,,}   &\vec{\theta}_R \in [0, 2\pi)\,. 
\phantom{\qquad   100\le m_{\nu_{D_3}}/{\text GeV}}&\phantom{<  500  }\nonumber
\end{alignat}
%\begin{eqnarray}
%&&1 \le m_{\nu_{D_1}}/{\text GeV}<  10\,, \qquad   10\le m_{\nu_{D_2}}/{\text GeV}<  100\,, 
%\qquad   100\le m_{\nu_{D_3}}/{\text GeV}<  500  \nonumber \\ && \\ \nonumber 
%&&10^{13}~{\text GeV} \le M_{R_1}/{\text GeV}<  10^{14}~{\text GeV}\,, \qquad   10^{14}\le M_{R_2}/{\text GeV}<  10^{15}\,, 
%\qquad   10^{15}\le M_{R_3}/{\text GeV}<  10^{16} \\ \nonumber \\
%&& \vec{\theta}_R \in [0, 2\pi)\,.\nonumber 
%\end{eqnarray}
Notice that, being the neutrino masses given by complicated expressions of parameters, the position $m_{\nu_{D_1}}<m_{\nu_{D_2}}<m_{\nu_{D_3}}$ and $M_{R_1}<M_{R_2}<M_{R_3}$
does not correspond a priori to a definite mass hierarchy, as it would be the case for a standard see-saw mechanisms where, for example, $m_i\sim m^2_{\nu_{D_i}}/M_{R_i}$ for NO.
We have analyzed the 6 different cases corresponding to modified BM, TBM and GR and the two values of $\sigma=\pm 1$; for each texture, we reported in Tab.(\ref{fit_table}) the minimum of $F(\vec{m}_{\nu_{D}},\vec{M}_{R},\vec{\theta}_R)$ and the values of the vectors $\vec{m}_{\nu_{D}},\vec{M}_{R}$ and $\vec{\theta}_R$ in which the minimum is assumed. We also report in Fig.(\ref{posterior}) an example of posterior distributions for the BM case, $\sigma=-1$ (all cases are very similar to each other).
\begin{table}[h]
\centering
\renewcommand{\arraystretch}{1.2}
\begin{tabular}{l c c c c c } 
\toprule
&  &$F^{\text min}$ & $\vec{m}_{\nu_{D}}$ (GeV) & $\vec{M}_{R}$ ($10^{13}$ GeV)& $\vec{\theta}_R$ ($^\circ$)\\ 
\hline
& $\sigma=+1$ & 0.42 & (9.35, 55.35, 117.70) & (4.0, 70.10, 297.51)& (145.80, 195.44, 162.01) \\
BM &$\sigma=-1$ & 0.44 & (9.98, 36.84, 111.52) & (3.59, 67.78, 708.16)& (122.88, 12.88, 112.59)\\ \hline
& $\sigma=+1$ & 0.31 & (9.38, 58.54, 130.56)   & (3.41, 93.21, 794.35) &  (331.00, 347.73, 317.70)\\
TBM &$\sigma=-1$ & 0.34  & (9.66, 34.52, 159.29)  & (2.09, 55.69, 485.30) & (321.58, 354.03, 221.83)  \\
\hline
& $\sigma=+1$ & 0.19 & (9.83, 80.88, 201.77) &(3.55, 86.014, 728.73) & (341.65, 171.70, 188.76)\\
GR &$\sigma=-1$ &  0.45 & (9.26, 42.48, 156.88) & (2.97, 38.95, 268.45) & (145.00, 8.15, 341.63) \\
\bottomrule
\end{tabular}
\caption{\it Results of the minimization procedure of the function $F(\vec{m}_{\nu_{D}},\vec{M}_{R},\vec{\theta}_R)$ in eq.(\ref{ffunction}). $F^{min}$ stands for the minimum value of such a function; the meaning of the three vectors 
$\vec{m}_{\nu_{D}}$, $\vec{M}_{R}$ and $\vec{\theta}_R$
has been given in eq.(\ref{vectors}).}
\label{fit_table}
\end{table}
Let us analyze more in detail the results of our minimizing procedure. 
%Looking at the posterior distributions of the nine observables in $\vec{m}_{\nu_{D}}, \vec{M}_{R}=(M_{R_1},M_{R_2},M_{R_3})$ and $\vec{\theta}_R$. 
First of all, none of the analyzed patterns can be tagged as a preferred one, as the minima of the $F$ function are very close to each other. This is in agreement with what we found for the mixing angles where, after including all relevant corrections,  no preferred choice emerged. 
The vector $\vec{m}_{\nu_{D}}$ is characterized by the fact that the first and third element prefer values at their upper and lower limits, respectively while $m_{\nu_{D_2}}$ is generally confined in the central region (with an exception for the case TBM, $\sigma=+1$ which, instead, prefers larger values). As for the Majorana masses, we observe similarities in all elements among the different patterns: $M_{R_1}$ and $M_{R_3}$ tend to stay close to their allowed lower and upper bounds, respectively,  while $M_{R_2}$ is mostly concentrated in the middle region around $[40-90]\cdot10^{13}$ GeV. It is interesting to observe that the posterior distributions (middle panels of Fig.(\ref{posterior})) are almost flat for  $M_{R_3}$  but peaked at large allowed values for $M_{R_1}$ and $M_{R_2}$; while for the latter case this seems consistent with the values at the minimum of $F$, for the former this behavior does not completely match what reported in Tab.(\ref{fit_table}). We interpret this as that $M_{R_1}$ gives a smaller contribution to $F$ as the other Majorana masses. This happens also for the first Dirac neutrino mass $m_{\nu_{D_1}}$, whose best fit value is close to its upper limit while the posterior distribution is essentially flat. Finally, a look at Fig.(\ref{posterior}) reveals that the posterior distributions for the mixing angles are multi-modal; in particular, a clear bi-modal distribution is seen for $\theta_{R_{12}}$, around $|\sin(\theta_{R_{12}})|\sim 1/2$, and for  
$\theta_{R_{13}}$ around $|\sin(\theta_{R_{13}})|\sim 0$; this is also visible in Tab.(\ref{fit_table}). A less clear bi-modal behaviour is also present for $\theta_{R_{23}}$ but the spreads around the maximum posterior probability are not negligible. 
Assuming the fixed values $\sin(\theta_{R_{12}})=1/2$ and  $\sin(\theta_{R_{13}})=0$, the right-handed rotation implied by our fit is as follows:
\begin{equation}
\label{uright}
U_{R} =
\begin{pmatrix}
\sqrt{3}/2 & 1/2 & 0\\
-c_{23}/2 & \sqrt{3} c_{23}/2
 & s_{23}\\
s_{23}/2 &
-\sqrt{3} s_{23}/2 & c_{23}
\end{pmatrix}
\,.
\end{equation}

\section{Conclusions} 
\label{conclusions}
In this paper we have investigated in detail the hypothesis that the PMNS mixing matrix is given by the relation $U_{PMNS} =V_{CKM}^\star\,T^\star$, where $T$ is a unitary matrix. By considering the decomposition
$T \equiv U_{23} U_{13} U_{12}$, we have shown that a $T$ matrix coinciding with TBM, BM and GR mixing fails, among others, to reproduce the experimental preferred value of the Jarlskog invariant, which is related to the third power of the Cabibbo angle. To solve these issues, we have analyzed ${\cal O}(\lambda)$ corrections to the $U_{ij}$ matrices, showing that a complex parameter $u$ is needed in the (13) rotation to reconcile our ansatz with the experimental amount of leptonic CP violation. While a correction $\omega$ in the (23) sector is needed for a substantial deviation  of the atmospheric angle from maximality, it only marginally improves the global fit to the experimental values of the mixing angles, because of a  wrong estimate of $\theta_{12}$ in all cases but BM. Thus, a shift in the (12) plane is mandatory to  account for the solar angle and, consequently, to get an excellent fit for all mixing parameters and for any initial choice of $T$.
The ansatz illustrated here is also appropriate to reproduce the value of solar and atmospheric mass differences. Indeed, equipped with the best fit values of the $\re(u), \im(u), \omega$ and $z$ parameters, we have shown that a description of neutrino masses via the see-saw mechanism is possible. Because of the cumbersome analytical expressions of $\Delta m^2_{sol,atm}$, we relied on numerical scan of the vector components of  $\vec{m}_{\nu_{D}}$, $\vec{M}_{R}$ and $\vec{\theta}_R$ of eq.(\ref{vectors}) and found that, with our choice of priors,  a complete description of neutrino masses and mixing under the assumption $U_{PMNS} =V_{CKM}^\star\,T^\star$ is possible.

\section*{Acknowledgments} 
We thank Jo\~ao Penedo and Matteo Parriciatu for useful comments and suggestions on our manuscript.

\section*{Appendix: Full ${\cal O}(\lambda^3)$ formulae} 
\label{appendix}
For the sake of completeness, we report here the full ${\cal O}(\lambda^3)$ expressions of the mixing parameters obtained from our ansatz  $U_{PMNS} =V_{CKM}^\star\,T^\star$. 
\begin{align}
&J=\frac{\lambda}{4} \,\sigma\,
\im{(u)} \, \sin(2 \tilde{\theta}_{12})+\lambda^2 \frac{\im(u)\cos^2(\tilde{\theta}_{12})}{4\cos(\tilde{\theta}_{12})}
\left(\sqrt{2}\cos(\tilde{\theta}_{12})  + 2 z\right)  + \nonumber \\
& -\frac{\lambda^3}{8}\sigma\sin(2 \tilde{\theta}_{12})
\left[\sqrt{2} A \eta + 2 \im(u) \left(2 + |u|^2 + \sqrt{2} \re(u)\right) \right]+\\
&-\frac{\lambda^3}{2} \,\im(u)\,\left[\omega \cos\left(2 \tilde{\theta}_{12}\right) + 2 \sigma \omega^2 \sin\left(2 \tilde{\theta}_{12} \right)\right]+\nonumber \\ 
&- \lambda^3 z \sigma\left\{-8 \sqrt{2} \sin(\tilde{\theta}_{12}) \im(u) +
\frac{2 z \im(u)\sin(\tilde{\theta}_{12})}{\cos(\tilde{\theta}_{12})^3}  \left[3 \cos^2(\tilde{\theta}_{12})  + \sin^2(\tilde{\theta}_{12})\right] \right\}
\label{jarls2}\nonumber
\end{align}

\begin{eqnarray}
\label{reactor2}
\sin(\theta_{13})&=&    \sqrt{1/2 + |u|^2 + \sqrt{2} \re(u)}\, \lambda+ \nonumber \\
&&+ \frac{\lambda^2 \omega}{\sqrt{2}} \,\frac{\left[\sqrt{2} + 2 \re(u)\right] }
 {\sqrt{1 + 2 |u|^2 + 2 \sqrt{2} \re(u)}}\nonumber +\\
&&+\lambda^3\,\frac{\left[2 \sqrt{2} A \rho - 4 A \eta \im{u} + (-2 + 4 A \rho) \re{u} - 
  |u|^2 (3 \sqrt{2} + 2 \re(u))\right]}{4\sqrt{1 + 2|u|^2 + 2\sqrt{2} \re(u)}}+\, \\
&&   +\lambda^3\,\omega^2\,\frac{\sqrt{2} \im^2(u)}{\left[2 \re(u) \left(\re(u)+\sqrt{2}\right)+2 \im^2(u)+1\right]^{3/2}}\nonumber
\end{eqnarray}

\begin{eqnarray}
\label{atm3}
\tan(\theta_{23})&=&   1 +2 \sqrt{2} \lambda \omega+\frac{\lambda^2}{2} \left[ -1 + 4 A-2 \sqrt{2} \re(u)\right]
+ \lambda ^3 \omega  \left(4 \sqrt{2} A-2 \re(u)+12 \sqrt{2} \omega ^2-\sqrt{2}\right)
\nonumber \\ \,.
\end{eqnarray}

\begin{eqnarray}
\label{sol3}
\tan(\theta_{12})&=&  \tan(\tilde{\theta}_{12}) + \frac{\lambda\,\sigma  \left(\sqrt{2} \tilde{c}_{12}+2 z\right)}{2 \tilde{c}_{12}^3}+
 \frac{\lambda^2}{2 \tilde{c}^3_{12}} \tilde{s}_{12} -\frac{\lambda^2 \omega \sigma}{\tilde{c}^2_{12}}
 +\frac{\lambda^2\,z\sqrt{2}\tilde{s}_{12}}{\tilde{c}^4_{12}}+\frac{3\lambda^2 z^2 \tilde{s}_{12}}{2\tilde{c}^5_{12}}+
 \nonumber \\
 &+& \frac{\lambda^3}{4\tilde{c}^4_{12}} \sigma\,
 \left[\sqrt{2} (1 - 2 A \tilde{c}_{12}^2 \rho) + \tilde{c}_{12}^2 (\sqrt{2} |u|^2 + 2 \re(u)) \right] + \\
\nonumber &+& \frac{\lambda^3}{2\tilde{c}^7_{12}}\left\{-2 \sqrt{2} \tilde{c}^4_{12} \omega  (\tilde{c}_{12} \sigma  \omega +\tilde{s}_{12})+\tilde{c}^2_{12} z \left[\tilde{c}^2_{12} \sigma -2
   \omega  \sin (2 \tilde{\theta}_{12})+3 \sigma  \tilde{s}^2_{12}\right]+ \right. \\
\nonumber &&  \left.\frac{\tilde{c}_{12}\sigma  z^2 (5-3 \cos (2
   \tilde{\theta}_{12}))}{\sqrt{2}}+\sigma  z^3 (3-2 \cos (2 \tilde{\theta}_{12}))\right\}\nonumber\,
\end{eqnarray}

\renewcommand\bibname{Bibliography}
\bibliographystyle{utphys}
\bibliography{references}{}

\end{document}